\documentclass[a4paper]{article}

\usepackage{graphicx}
\usepackage{float}
\usepackage{amsmath}

\title{Profiled spectral lines generated by Keplerian discs orbiting in the Bardeen and Ayon-Beato-Garcia spacetimes}
\author{Jan Schee and Zden\v{e}k Stuchl\'{i}k\\
 \small{Institute of Physics and Research Centre for Theoretical Physics \& Astrophysics}\\
        \small{Faculty of Philosophy and Science}\\
        \small{Silesian university in Opava}\\
        \small{Bezru\v{c}ovo n\'{a}m. 13, 74601 Opava}\\
        \small{Czech Republic}
}

\date{}

\newcommand{\beq}{\begin{equation}}
\newcommand{\eeq}{\end{equation}}
\newcommand{\bea}{\begin{eqnarray}}
\newcommand{\eea}{\end{eqnarray}}
\newcommand{\diff}{\mathrm{d}}

\begin{document}
\maketitle
\abstract{Shape and frequency-shift map of direct and indirect images of Keplerian discs orbiting in Bardeen and Ayon-Beato-Garcia (ABG) black hole and no-horizon spacetimes are determined. Then profiles of spectral lines generated in the innermost parts of the Keplerian discs in the Bardeen and ABG spacetimes are constructed. The frequency-shift maps and profiled spectral lines are compared to those generated in the field of Schwarzschild black holes and possible observationally relevant signatures of the regular black hole and no-horizon spacetimes are discussed. We demonstrate that differences relative to the Schwarzschild spacetimes are for the no-horizon spacetimes much more profound in comparison to the regular black hole spacetimes and increase with increasing charge parameter of the spacetime. The differences are stronger for small and large inclination angles than for mediate ones. For the no-horizon spacetimes, the differences enable to distinguish the Bardeen and ABG spacetime, if inclination angle to the distant observer is known. We also show that contribution of the so called ghost images to the profiled lines increases with increasing charge parameter of the spacetimes.}

\section*{Introduction}
The black holes governed by standard general relativity contain a physical singularity with diverging Riemann tensor components and predictability breakdown, considered as a realm of quantum gravity overcoming this internal defect of general relativity. However, families of regular black hole solutions of Einstein's gravity have been found that eliminate the physical singularity from the spacetimes having an event horizon. Of course, these are not vacuum solutions of the Einstein equations, but contain necessarily a properly chosen additional field, or modified gravity, so that the energy conditions related to the existence of physical singularities \cite{Haw-Elli:1973:LargeScaleStructure:} are then violated. 

The regular black hole solution containing a magnetic charge as a source parameter has been proposed by Bardeen \cite{Bar:1968:GR5Tbilisi:}, having the magnetic charge related to a non-linear electrodynamics \cite{AyB-Gar:2000:PhysLetB:}. The other solution of the combined Einstein and non-linear electrodynamic equations has been introduced by Ayon-Beato and Garcia \cite{AyB-Gar:1998:PhysRevLet:,AyB-Gar:1999:PhysLetB:,AyB-Gar:1999:GenRelGrav:}. Both these solutions are characterized by the mass parameter $m$ and the charge parameter $g$. Their geodesic structure is governed by the dimensionless ratio $g/m$ giving the specific charge of the source. Details of the properties of the electromagnetic fields related to the spherically symmetric Bardeen and Ayon-Beato-Garcia (ABG) spacetimes were discussed by Bronnikov \cite{Bro:2000:PHYSRL:,Bro:2001:PHYSR4:}. A different approach to the regular black hole solutions was applied by Hayward \cite{Hay:2006:PhysRevLet:}. Modification of the mass function in the Bardeen and Hayward solutions and inclusion of the cosmological constant can be found in the new solutions of Neves and Saa \cite{Nev-Saa:2014:arXiv:1402.2694:}. Rotating regular black hole solutions were introduced in \cite{Mod-Nic:2010:PHYSR4:,Bam-Mod:2013:PhysLet:,Tos-Abd-Ahm-Stu:2014:PHYSR4:,Azr:2014:PHYSR4:}. 

For properly chosen charge parameter $g/m$, the Bardeen and ABG solutions allow for existence of fully regular spacetime, without an event horizon. We call such solutions "no-horizon" spacetimes. Some of their properties were discussed in \cite{Pat-Jos:2012:PHYSR4:}. We shall consider here both the black hole and no-horizon spacetimes. 

Dynamics of test particles and fields around regular black holes has been recently discussed in \cite{Pat-Jos:2012:PHYSR4:,Tos-Abd-Ahm-Stu:2014:PHYSR4:,Azr:2014:PHYSR4:,Gho-She-Ami:2014:PHYSR4:,Tos-Abd-Ahm-Stu:2015:ASS:,Tos-Abd-Stu-Ahm:2015:PHYSR4:,Gar-Hac-Kun-Lam:2015:JMP:,Gho-Mah:2015:EJPC:,Mac-Oli-Cri:2015:PHYSR4:}. A detailed discussion of the geodesic structure of the regular Bardeen and ABG black hole and no-horizon spacetimes and its implication to simple optical phenomena as the silhouette shape and extension and the profiled spectral lines generated by Keplerian rings constituted from test particles following stable circular geodesics were presented in \cite{Stu-Sche:2015:IJMPD:}. It has been demonstrated that the geodesic structure of the regular black holes outside the horizon is similar to those of the Schwarzschild or Reissner-N\"{o}rdstr\"{o}m (RN) black hole spacetimes, but under the inner horizon, no circular geodesics can exist. The geodesic structure of the no-horizon spacetimes is similar to those of the naked singularity spacetimes of the RN type \cite{Stu-Hle:2002:ActaPhysSlov:,Pug-Que-Ruf:2011:PHYSR4:} related to the Einstein gravity, or the Kehagias-Sfetsos (KS) type \cite{Stu-Sche:2014:CLAQG:,Stu-Sche-Abd:2014:PHYSR4:} that is related to the solution of the modified Ho\v{r}ava quantum gravity \cite{Hor:2009:PHYSR4:,Hor:2009:PHYSRL:} in the infrared limit \cite{Keh-Sfe:2009:PhysLetB:}. In all of these no-horizon and naked singularity spacetimes, an "antigravity" sphere exists consisting of static particles located at stable equilibrium points at a given "static" radius that can be surrounded by a Keplerian disc. 

In the naked singularity and no-horizon spacetimes, additional optical images of Keplerian discs in comparison to those related to the images observed in the field of black holes exist \cite{Stu-Sche:2010:CLAQG:,Stu-Sche:2014:CLAQG:,Sche-Stu:2015:JCAP:}. They occur in the region that should be empty in the case of images created in the black hole spacetimes -- for this reason we call them ghost images. Moreover, character of the ghost images differs in the case of the naked singularity spacetimes, and the no-horizon spacetimes, because of different character of the motion of photons having small values of the impact parameter. Such photons feel strong repulsion in the naked singularity spacetimes, but they feel very weak influence in the central region of the no-horizon spacetimes \cite{Sche-Stu:2015:JCAP:}. 

Properties of the Keplerian discs in the Bardeen and ABG no-horizon spacetimes strongly depend on the value of the charge parameter. If the charge parameter $g/m$ is close to value corresponding to the extreme black-hole state, two Keplerian discs exist above the static radius, and even two photon circular orbits exist in the near-extreme states, the inner one being stable representing the outer edge of the inner Keplerian disc, while the outer disc is limited by the innermost stable circular geodesic. As an exceptional phenomenon, not occuring in the naked singularity or the Bardeen no-horizon spacetimes, an internal Keplerian disc can occur under the antigravity sphere in the ABG no-horizon spacetimes with $g/m > 2$ \cite{Stu-Sche:2015:IJMPD:}. Such an internal disc can be visible, if matter is not filling whole the static radius. 

It is important to look for observationally relevant phenomena of the regular Bardeen and ABG black-hole and no-horizon spacetimes. In strong gravity, three observationally important tests are widely discussed -- the spectral continuum \cite{Rem-McCli:2006:ARAA:,McCli-etal:2011:CLAQG:}, the profiled spectral lines \cite{Laor:1989:IUAS:,Bao-Stu:1992:ApJ:,Bao-Had-Oest:1996:ApJ:,Fan-etal:1997:PASJ:,Sche-Stu:2009:GenRelGrav:,Sche-Stu:2013:JCAP:}, and high-frequency quasiperiodic oscillations \cite{Stu-Kot:2009:GenRelGrav:,Stu-Sche:2012a:CLAQG:}. Here we calculate the profiled spectral lines generated at the innermost parts of the Keplerian accretion discs related to stable circular geodesics \cite{Nov-Tho:1973:BlaHol:}. The profiled spectral lines generated at the spherically symmetric regular black-hole and no-horizon spacetimes are compared to those generated under corresponding conditions around Schwarzschild black holes. 

First, we compare for the Bardeen and ABG black hole and no-horizon spacetimes the direct and indirect images, including the ghost images, of the innermost parts of the Keplerian discs radiating in a fixed frequency, giving the shape of the images and map of the frequency shift of the radiation. Then we calculate profiled spectral lines generated by the innermost parts of the Keplerian discs taking into account the condition of the negative gradient of angular velocity in the Keplerian discs \cite{Stu-Sche:2015:IJMPD:} that is necessary condition for the accretion governed by the magnetorotational instability (MRI) \cite{Bal-Haw:1991:,Bal-Haw:1998:RevModPhys:}, as discussed in \cite{Stu-Sche:2014:CLAQG:,Vie-etal:2014:PHYSR4:}. We also demonstrate the role of the ghost images in the shaping of profiled spectral lines. 

The electromagnetic field related to the spherically symmetric regular Bardeen and ABG black hole (or no-horizon) spacetimes is discussed in \cite{AyB-Gar:1998:PhysRevLet:,AyB-Gar:1999:PhysLetB:,AyB-Gar:1999:GenRelGrav:}. Due to the electrodynamics non-linearity, the motion of photons could deviate from the null geodesics of the spacetimes. However, here we do not consider these deviations keeping the assumption of the photon motion governed by the null geodesics of the spacetime, as in related previous studies. Then the electromagnetic field is irrelevant for our study and we consider the geometry properties only, both for the Keplerian (circular) geodesic motion assumed for accretion discs, and the null geodesics of the spacetime related to the photon motion. We assume that at the centre of coordinates, $r=0$, the self-gravitating source of the electromagnetic field of the background is located, where trajectories of test particles and photons terminate, similarly to the case of the central singular points in the spherically symmetric naked singularity spacetimes \cite{Sche-Stu:2015:JCAP:}.

\section{Regular Bardeen and Ayon-Beato-Garcia spacetimes}
The spherically symmetric geometry of the regular Bardeen and ABG black-hole or no-horizon spacetimes is characterized in the standard spherical coordinates and the geometric units (c=G=1) by the line element
\beq
	\diff s^2=-f(r)\diff t^2+\frac{1}{f(r)}\diff t^2+r^2(\diff\theta^2+\sin^2\theta\diff\phi^2),\label{interval}
\eeq
where the "lapse" $f(r)$ function depends only on the radial coordinate, the gravitational mass parameter $m$. and the charge parameter $g$. Both Bardeen and ABG spacetimes are constructed to be regular everywhere, i.e., the components of the Riemann tensor, and the Ricci scalar are finite at all $r\ge 0$ \cite{AyB-Gar:1999:GenRelGrav:}.

The lapse function $f(r)$ reads 
\begin{enumerate}
\item Bardeen spacetime
\beq
	f(r)=1 - \frac{2 m r^2}{(g^2 + r^2)^{3/2}}.
\eeq
\item ABG spacetime
\beq
	f(r)=	1  - \frac{2 m r^2}{(g^2 + r^2)^{3/2}} + \frac{g^2 r^2}{(g^2 + r^2)^2}.
\eeq

\end{enumerate}

The event horizons of the Bardeen and ABG spacetimes are given by the condition  
\beq
f(r)=0. \label{pseudosingularity}
\eeq 
The loci of the black hole horizons are determined by the relations  
\begin{enumerate}
\item Bardeen 
	\beq
		g^6 + (3g^2 - 4m^2)r^4 + 3g^4r^2 + r^6 = 0 ;
	\eeq
\item ABG 
	\beq
	    (g^2 + r^2)^2(g^4 + 4g^2r^2 + r^4) - 4m^2r^4(g^2 + r^2) + g^4r^4 = 0 .
	\eeq
\end{enumerate} 
The solutions for the location of the event horizons are presented in Figure 1. If real and positive solutions of the equation (\ref{pseudosingularity}) do not exist, the spacetime is fully regular, having no event horizon. We call it "no-horizon" spacetime. The critical values of the dimensionless parameter $g/m$ separating the black-hole and the "no-horizon" Bardeen and ABG spacetimes read 
\begin{enumerate}
\item Bardeen 
	\beq
		(g/m)_{NoH/B} = 0.7698 ;
	\eeq
\item ABG 
	\beq
	    (g/m)_{NoH/ABG} = 0.6342 .
	\eeq
\end{enumerate} 
In the "no horizon" Bardeen and ABG spacetimes the metric is regular at all radii $r \geq 0$. We assume $r=0$ to be the site of the self-gravitating charged source of the spacetime; test particle or photon trajectories terminate at $r=0$. For more details on the properties of the Bardeen and ABG spacetimes see \cite{Stu-Sche:2015:IJMPD:,Sche-Stu:2015:JCAP:}. 

\section{Keplerian discs and classification of the Bardeen and ABG spacetimes}

The Keplerian accretion discs are governed by circular geodesics of the background spacetime. In the spherically symmetric spacetimes the geodesics are tied to central planes. For the Keplerian discs we choose the central plane to be the equatorial plane $\theta = \pi/2$. The spacetime symmetries imply existence of two constant of motion, the energy $E$ related to the timelike Killing vector, and the axial angular momentum $L$ related to the axial Killing vector that coincides with the total angular momentum for the motion in the equatorial plane. The rest mass-energy $\mu > 0$ is the other constant of motion. Then it is convenient to consider the specific energy $E/\mu$ and specific angular momentum $L/\mu$. For simplicity we can put $\mu = 1$. 

\subsection{Effective potential and circular geodesics}
The circular motion of test particles is determined by an effective potential related to the specific angular momentum $L$ related to the particle rest mass $\mu$ that takes in the equatorial plane of spherically symmetric spacetimes a simple form
\beq
            V_{eff} = f(r)\left(1+\frac{L^2}{r^2}\right). 
\eeq
The radial equatorial motion of a particle with specific energy $E$ is then governed by the equation for the radial component of the 4-velocity 
\beq
           (u^r)^2 = E^2 - V_{eff} . 
\eeq
For the photon motion, $\mu = 0$, the effective potential is related to the impact parameter $b=E/L$ and reads  
\beq
           V_{ph} = f(r)\left(\frac{b^2}{r^2}\right). 
\eeq
The properties of the effective potential outside the event horizon are similar to those in the Schwarzschild spacetimes, while in the no-horizon spacetimes they are similar to the case of RN naked singularity spacetimes, and are discussed in \cite{Stu-Sche:2015:IJMPD:}. 

The specific angular momentum $L_c$, the specific covariant energy $E_c$, and the angular frequency relative to distant observer $\Omega_c$ of the circular geodesic orbits at a given radius $r$ in the Bardeen and ABG spacetimes take the form \cite{Stu-Sche:2015:IJMPD:}
\begin{enumerate}
\item \emph{Bardeen}
\bea
	L^2_c&=&\frac{m r^4 (2 g^2 - r^2)}{3 m r^4 - (r^2 + g^2)^{5/2}},\\
	E^2_c&=&\frac{\left[-2 m r^2 + (r^2 + g^2)^{3/2}\right]^2}{g^6 + 3 g^4 r^2 + 3 g^2 r^4 + r^6 - 
	 3 m r^4 \sqrt{r^2 + g^2}}\\
	 \Omega^2_c&=&\frac{m(r^2-2g^2)}{(r^2+g^2)^{5/2}}
\eea 

\item \emph{ABG}
\bea
	L^2_c&=&\frac{(r^4 (g^2 (g^2 - r^2) \sqrt{g^2 + r^2} + 
	   m (-2 g^4 - g^2 r^2 + r^4)))}{(-3 m r^4 (g^2 + r^2) + 
	 \sqrt{g^2 + r^2} (g^6 + 3 g^4 r^2 + 5 g^2 r^4 + r^6))},\\
	E^2_c&=&\frac{(g^4 \sqrt{g^2 + r^2} + r^4 (-2 m + \sqrt{g^2 + r^2}) + 
	  g^2 r^2 (-2 m + 3 \sqrt{g^2 + r^2}))^2}{((g^2 + r^2)^2 (g^6 + 
	   3 g^4 r^2 + 5 g^2 r^4 + r^6 - 3 m r^4 \sqrt{g^2 + r^2}))},\\
	\Omega^2_c&=&\frac{m r^4 + g^4 (-2 m + \sqrt{r^2 + g^2}) - g^2 r^2 (m + \sqrt{g^2 + r^2})}{(r^2+g^2)^{7/2}}
\eea 
\end{enumerate} 

\begin{figure}[H]
\begin{center}
\begin{tabular}{cc}
\includegraphics[scale=0.6]{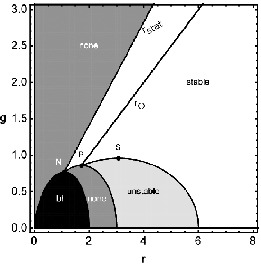}&\includegraphics[scale=0.6]{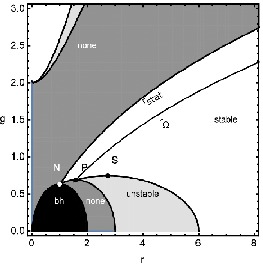}
\end{tabular}
\caption{Loci of circular orbits in Bardeen (left) and ABG (right) spacetimes. The white regions correspond to stable orbits, while the light gray regions represent unstable orbits. In the dark gray regions no circular orbits exist. In the Bardeen case the characteristic values of the specific charge $g$ are given by the points $S=(3.07,0.95629)$, $N=(1.09,0.76988)$, and $P=(1.7179, 0.85865)$. In the ABG case the characteristic values of the specific charge $g$ are given by the points $S=(2.73,0.74684)$, $N=(0.9941,0.6343)$, and $P=(1.57, 0.690771)$. The loci of the horizons are given by the boundary of the black region. The black line represents loci of the photon circular orbits. The curves $r_{\Omega}(g)$ terminated at the point $P$ determine the loci of maximal angular velocity of the circular motion, giving a natural boundary of Keplerian discs in the no-horizon spacetimes. \label{fig1}}
\end{center}
\end{figure}



We now briefly summarize properties of the circular geodesics that were in detail presented in \cite{Stu-Sche:2015:IJMPD:}. First we give two significant radii occuring in all the Bardeen and ABG no-horizon spacetimes.

\subsection{Antigravity sphere}

In all the Bardeen and ABG no-horizon spacetimes a static radius exists where an "antigravity" effect of the geometry is demonstrated by vanishing of the specific angular momentum, $L_c = 0$, similarly to the case of the KS naked singularity spacetimes of the modified Ho\v{r}ava gravity \cite{Stu-Sche:2014:CLAQG:,Vie-etal:2014:PHYSR4:}, or to the case of the RN naked sigularity spacetimes \cite{Stu-Hle:2002:ActaPhysSlov:,Pug-Que-Ruf:2011:PHYSR4:}. The static radius is given by 
\begin{enumerate}
\item \emph{Bardeen} 
\beq
	r_{stat} = \sqrt{2} g
\eeq
\item \emph{ABG} 
\beq
	g^2 (g^2 - r^2) \sqrt{g^2 + r^2} +  m (-2 g^4 - g^2 r^2 + r^4) = 0 . 
\eeq
\end{enumerate}
Location of the static radius in dependence on the charge parameter $g/m$ of the Bardeen and ABG no-horizon spacetimes is demonstrated in Figure 1. The static radius corresponds to an "antigravity" sphere where the particles can remain is stable static equilibrium. The antigravity sphere, being surrounded by a Keplerian disc, can be considered as a final state of the accretion process, and in some sense can represent the effective surface of the objects described by the Bardeen of ABG no-horizon spacetimes. No circular orbits are possible under the stable static radius, except the case of the ABG no-horizon spacetimes with the charge parameter $g/m > 2$ when "internal" circular orbits can exist under the stable static radius \cite{Stu-Sche:2015:IJMPD:}. The inner part of the ABG internal circular geodesics starting at $r=0$ is stable, and is separated by an outermost stable circular geodesic from the region of unstable circular geodesics terminating at an outer edge of the circular geodesics given by a secondary static radius corresponding to static particles in an unstable static equilibrium position \cite{Stu-Sche:2015:IJMPD:} -- see Figure 1. 


\subsection{Radius of local maximum of the angular frequency of circular geodesics}

The standard Keplerian accretion requires increasing gradient of the angular frequency of the orbiting matter, $\diff\Omega^2_c/\diff r < 0$, if the MRI accretion mechanism should be at work \cite{Bal-Haw:1991:}. However, this condition is not fulfilled for all stable circular geodesics in the no-horizon Bardeen and ABG spacetimes, similarly to the spherically symmetric naked singularity spacetimes of the KS type \cite{Stu-Sche:2014:CLAQG:}. 

The local maximum of the function $\Omega_c^2(r;g,m)$ is located along the curve $r_{\Omega}(g,m)$ determined by the condition $\diff\Omega^2_c/\diff r=0$ that for the particular spacetime imply the formulae 
\begin{enumerate}
\item \emph{Bardeen}
	\beq
		r_{\Omega/B} = 2g;
	\eeq
\item \emph{ABG} 
where $r_{\Omega/ABG}$ is determined in an implicit form by the relation 
	\beq
		-3 m r^5 + 4 g^4 r (3 m - 2 \sqrt{g^2 + r^2}) + 
		 g^2 r^3 (9 m + 4 \sqrt{g^2 + r^2})=0.
	\eeq
\end{enumerate}
The functions $r_{\Omega}(g)$ are given for both the Bardeen and ABG spacetimes in Figure 1. The radius $r_{\Omega}(g)$ can be considered as the inner edge of the standard Keplerian discs. Possible scenarios of the subsequent accretion, under the radius $r_{\Omega}(g)$, are discussed in \cite{Stu-Sche:2014:CLAQG:} and will not be repeated here. 

\subsection{Classification of the Bardeen and ABG spacetimes according to circular geodesics} 

Now, we summarize the classification of the Bardeen and ABG spacetimes introduced in \cite{Stu-Sche:2015:IJMPD:}. In the no-horizon spacetimes, the classification is related to the character of the circular geodesics above the antigravity sphere. The classification is related to the existence of the photon circular geodesics (stable at $r{ph-s}$ and unstable at $r_{ph-u}$) and the marginally stable geodesics (giving the outer edge at $r_{OSCO}$ and the inner edge at $r_{ISCO}$) that are determined in \cite{Stu-Sche:2015:IJMPD:}. Dependence of these radii on the dimensionless charge parameter $g/m$ is given in Figure 1. The critical values of the charge parameter governing the spacetimes allowing for existence of photon (marginally stable) orbits $(g/m)_P$ ($(g/m)_S$) are given in Table 1. 

\begin{table}[H]
\begin{center}
\caption{Critical values of the dimensionless charge parameter $(g/m)$ of the Bardeen and ABG spacetimes.}
\begin{tabular}{|c|ccc|}
	\hline
	Charge parameter& $(g/m)_{NoH}$ & $(g/m)_P$ & $(g/m)_S$\\
	\hline
	Bardeen & $0.76988$ & $0.85865$ & $0.95629$ \\
	ABG & $0.6343$ & $0.690771$ & $0.74684$\\
	\hline
\end{tabular}
\end{center}
\end{table}

The classification of the Bardeen and ABG spacetimes according to the properties of the circular geodesics (at $r > r_{stat}$ in the no-horizon spacetimes) was given in \cite{Stu-Sche:2015:IJMPD:} and is summarized in Table 2; the classification holds equally for the Bardeen and ABG spacetimes. 

\begin{table}[H]
	\begin{center}
		\caption{Definition of the regular Bardeen and ABG spacetime classes. We denote Class I -  Black holes, Class II -  No-horizon with photon circular orbits, Class III - No-horizon with marginally stable circular orbits, Class IV - No-horizon allowing only stable circular orbits.}
		\begin{tabular}{|c|c|c|}
		\hline
			& Range of $g/m$ & Regions of circular geodesics\\
		\hline
		Class I & $0\leq g/m \leq (g/m)_N$ & $r_{ph}<r_{ISCO}<\infty$\\
		Class II & $(g/m)_N<g/m\leq (g/m)_P$ & $r_{stst}\leq r_{ph} \bigcup r_{ph}<r_{ISCO}<\infty$\\
		Class III  & $(g/m)_P<g/m\leq (g/m)_S$ & $r_{stat}<r_{\Omega}<r_{OSCO}<r_{ISCO}<\infty$\\
		Class IV  & $(g/m)_S<g/m$ & $r_{stat}<r_\Omega<\infty$\\
		\hline
		\end{tabular}
	\end{center}
\end{table}

In the Class II spacetimes, two regions of circular geodesics exist above the stable static radius. The outer one ranges, as in the black-hole spacetimes, from infinity down to the unstable photon circular geodesic.  The stable circular orbits exist down to the ISCO, under which unstable circular orbits exist being limited from below by the unstable photon circular orbit. The inner region consists of stable circular geodesics that range from the stable photon circular geodesic down to the static radius with stable equilibrium positions of test particles. It should be stressed that the inner region of stable circular geodesics is very extraordinary one, being very narrow, with the specific energy and specific angular momentum of the freely orbiting matter decreasing extremely steeply from arbitrarily high values down to the minimum at the stable static radius \cite{Stu-Sche:2015:IJMPD:}. 

In the Class III spacetimes, the circular geodesics extend from infinity down to the stable static radius. Two regions of stable circular geodesics are separated by a region of unstable circular geodesics. The outer region of stable orbits extends from infinity down to the ISCO, the inner region of the stable orbits extends from the OSCO down to the stable static radius. \footnote{Notice that $r_{OSCO}<r_{ISCO}$; $r_{OSCO}$ corresponds to the outer edge of the inner region of the stable orbits, while $r_{ISCO}$ denotes the inner edge of the outer region of the stable orbits.} In the regions of stable circular geodesics both the specific energy and specific angular momentum decrease with decreasing radius. The specific energy at the OSCO can be very high, if $g/m$ is close but above $(g/m)_{P}$ \cite{Stu-Sche:2015:IJMPD:}.

\section{Photon motion and direct and indirect images of Keplerian discs}

In \cite{Sche-Stu:2015:JCAP:} we have constructed the direct and indirect Keplerian disc images in the Bardeen spacetimes, with focus on the special case of the direct ghost images.  Here, we systematically compare appearance of the innermost parts of the Keplerian discs orbiting the regular Bardeen and ABG black hole, and no-horizon spacetimes of all three classes, and give their distinction to the images created by the Schwarzschild  black hole.

We present the optical appearance of both direct and indirect images of the Keplerian discs, reflecting their shape distortions due to gravitational lensing, and the frequency shift of the emitted radiation due to the gravitational and Doppler effects. We include also the ghost images related to the direct and indirect images, not going into details of their construction that have been exhaustively discussed in the case of the regular Bardeen no-horizon spacetimes and the Reissner-Nordstrom naked singularity spacetimes in \cite{Sche-Stu:2015:JCAP:}. We demonstrate the combined gravitational and Doppler shifts by a simple map assuming the Keplerian discs radiating locally at a fixed frequency corresponding, e.g., to a Fe X-ray line. In the following section, we construct profile of the spectral lines (usually assumed to be the fluorescent spectral Fe lines) generated in the innermost regions of the Keplerian discs, extending thus the results of the previous paper \cite{Stu-Sche:2015:IJMPD:} where profiled spectral lines were constructed in the simple case of radiating Keplerian rings near the ISCO. 

The disc appearance can be relevant for sources close enough when the current observational technique enables a detailed study of the innermost parts of the accretion structures located at vicinity of the black hole horizon or at the innermost parts of the no-horizon spacetimes. Such extremely precise observations are expected during few next years for the Sgr A*  source \cite{Doe-etal:2009:APJ:,Gwinn-etal:2014:ApJ:, Psaltis-etal:2014:ArXiv:, Ric-Dex:2015:MNRAS:,Psaltis-etal:2015:ApJ:,Chri-Loe:2015:PhRvD:}. On the other hand, the profiled spectral lines can be measurable also for very distant sources. 

\subsection{Photon motion}

For a general motion not confined to the equatorial plane where the Keplerian disc location is assumed, the trajectories of photons are independent of energy and can be related to  impact parameters 
\begin{equation}
                l = \frac{L}{E} , q = \frac{Q}{E},
\end{equation}
where $L$ is the axial angular momentum and $Q$ represents the total angular momentum \cite{Stu-Sche:2015:IJMPD:}.
For the non-equatorial photon motion, it is convenient to use the coordinates 
\begin{equation}
       u=\frac{1}{r} ,
\end{equation}
\begin{equation}
       m=\cos\theta .
\end{equation}
The equations of the radial and the latitudinal motion then take the form \cite{Stu-Sche:2014:CLAQG:}
\begin{equation}
\frac{du}{dw}=\pm u^{2}\sqrt{1-\widetilde{f}(u)\left(l^{2}+q\right)u^{2}}
\end{equation}
where  
\begin{equation}
\tilde{f}(u)=f(1/u) , 
\end{equation}
and
\begin{equation}
\frac{dm}{dw}=\pm u^{2}\sqrt{q-(l^{2}+q)m^{2}} 
\end{equation}
that can be properly integrated when photons radiated by Keplerian discs are considered \cite{Rau-Bla:1994:ApJ:,Sche-Stu:2009:IJMPD:,Sche-Stu:2009:GenRelGrav:,Stu-Sche:2010:CLAQG:}. 

\subsection{Frequency shift}

We assume the Keplerian discs to be composed from isotropically radiating particles following the circular geodesics. The frequency shift of radiation emitted by a point source moving along a circular orbit in the equatorial plane is given in the standard manner -- the frequency shift $\mathcal{G}$ between the emitter (e) and observer (o) is defined by the formula
\begin{equation}
\mathcal{G}=\frac{k_{\mu}U^{\mu}|_{o}}{k_{\mu}U^{\mu}|_{e}}.
\end{equation}
For circular orbits, the emitter four-velocity has only the time and axial components 
\begin{equation}
U^{\mu}=\left[U^{t},0,0,U^{\phi}\right].
\end{equation}
For the static observers at infinity, the frequency shift formula then reads
\begin{equation}
\mathcal{G}=\frac{1}{U_{e}^{t}(1-l\Omega)}, 
\end{equation}
where $l$ is the impact parameter of the photon, and $\Omega$ is the angular velocity of the emitter, here assumed to be the Keplerian angular velocity.  

\subsection{Apppearance of Keplerian discs}

Appearance of the Keplerian discs gives a basic physical information on the strong gravity influencing astrophysical processes around compact objects \cite{Sche-Stu:2009:IJMPD:,Stu-Sche:2010:CLAQG:,Abd-Ahm-Hak:2011:PHYSR4:,Stu-Sche:2012a:CLAQG:,Ata-Abd-Ahm:2013:ApSS:,Abd-Ata-Kuc-Ahm-Can:2013:ApSS:}. We shall construct the direct and indirect images of the Keplerian discs for small, mediate, and large inclination angle of the discs relative to the distant observers. In the case of regular Bardeen and ABG black holes, the Keplerian accretion discs have their edge at the ISCO. In the case of the no-horizon spacetimes allowing stable circular geodesics only, the Keplerian discs related to the MRI viscosity mechanism are considered to be physically relevant and will be studied here -- a detailed discussion can be found in the case of the Kehagias-Sfetsos naked singularity spacetimes in \cite{Stu-Sche:2014:CLAQG:}. Therefore, in the no-horizon spacetimes with $g/m < (g/m)_{S}$, we restrict our attention to the outer Keplerian discs located at $r > r_{ISCO}$, while in the spacetimes with $g/m > (g/m)_{S}$, we consider the part of the Keplerian discs at $r > r_{\Omega}$. We also demonstrate the possible role of the Keplerian disc located at $r_{stat}<r<r_{\Omega}$ for the Bardeen spacetime. \footnote{We do not consider the inner Keplerian discs in the case of $g/m < (g/m)_{S}$ spacetimes as existence of such discs is improbable from the astrophysical point of view \cite{Stu-Sche:2014:CLAQG:}.}

We are comparing the role of the specific charge of the spacetime in the appearance of the Keplerian discs. Since the regions of the specific charges related to the Bardeen and ABG no-horizon spacetimes of the same class are not overlapping in the case of the Classes II and III, admitting the unstable circular geodesics and the photon circular geodesics, we choose the values of the specific charge in the following way: 
\beq
      g_{II} = \frac{1}{2}(g_{NoH}+g_{P}) , g_{III} = \frac{1}{2}(g_{P}+g_{S}) . 
\eeq
In the case of the black hole spacetimes, and the no-horizon spacetimes with $g/m > (g/m)_{S}$, the value of the specific charge is chosen the same for the Bardeen and ABG spacetimes. 

We construct the direct and indirect images for characteristic values of the spacetime charge parameter $g/m$ (that will be for short denoted as $g$, assuming m=1) covering all four spacetime classes. The images are given for typical inclination angles of the Keplerian disc to the distant static observer $\theta_{o} = 30^{\circ}$, $60^{\circ}$, $85^{\circ}$. For each of the inclination angles, we first compare the Bardeen and ABG black hole images to the image created by the Schwarzchild black hole and then compare the Bardeen and ABG no-horizon Class II - IV spacetime images.
  
  The direct images are presented in Figures 2-7, while  indirect images are presented in Figures 8-13. The role of the region of the Keplerian disc where $\diff\Omega/\diff r > 0$, and the MRI condition is not satisfied, is demonstrated in Figs 14 and 15. In all the cases the outer edge of the Keplerian disc is located at $r_{out}=20m$. The images are expressed in terms of the standard coordinates $\alpha$ and $\beta$ introduced in the basic paper of Bardeen \cite{Bar:1973:BlaHol:}. The frequency shift $\mathcal{G}$ is represented by a colour code, and by few lines of $\mathcal{G}=const$. The frequency range reflecting the edges of the color code is explicitly given for each of the individual images. 


\begin{figure}[H]
	\begin{center}
		\begin{tabular}{c}
			\includegraphics[scale=0.4]{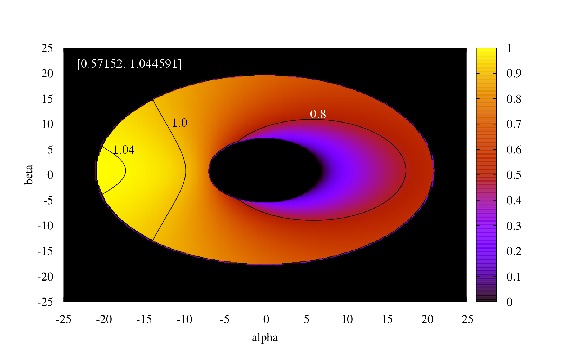}\\
			\includegraphics[scale=0.8]{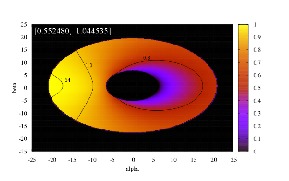}\\
			\includegraphics[scale=0.8]{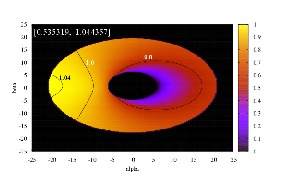}							
		\end{tabular}
		\caption{Frequency shift map on the Keplerian disc primary image. Images are constructed for Schwarzschild (top), Bardeen (middle) and ABG (bottom) black holes. The value of specific charge parameter is $g=0.5$ for both the Bardeen and ABG spacetimes.The observer inclination is $30^\circ$. The inner edge is at $r_{ISCO}$. The rotation of  disc is assumed anticlockwise here and in the following image figures. We give here and in the following image figures some lines of constant frequency shift; in brackets the range of the frequency shift is given.}
	\end{center}
\end{figure}

\begin{figure}[H]
\begin{center}
\begin{tabular}{cc}
	\includegraphics[scale=0.6]{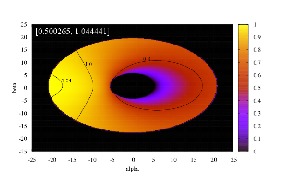}&\includegraphics[scale=0.6]{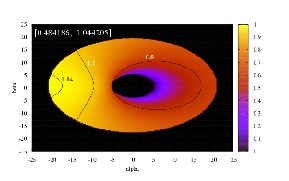}\\
	\includegraphics[scale=0.6]{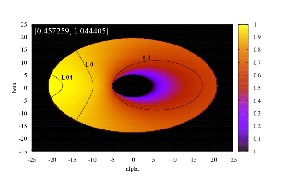}&\includegraphics[scale=0.6]{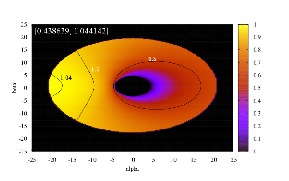}\\	
	\includegraphics[scale=0.22]{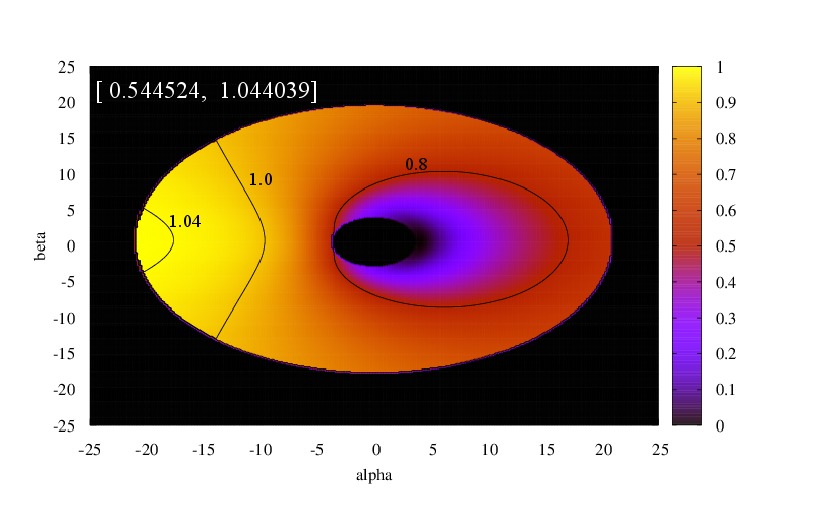}&\includegraphics[scale=0.22]{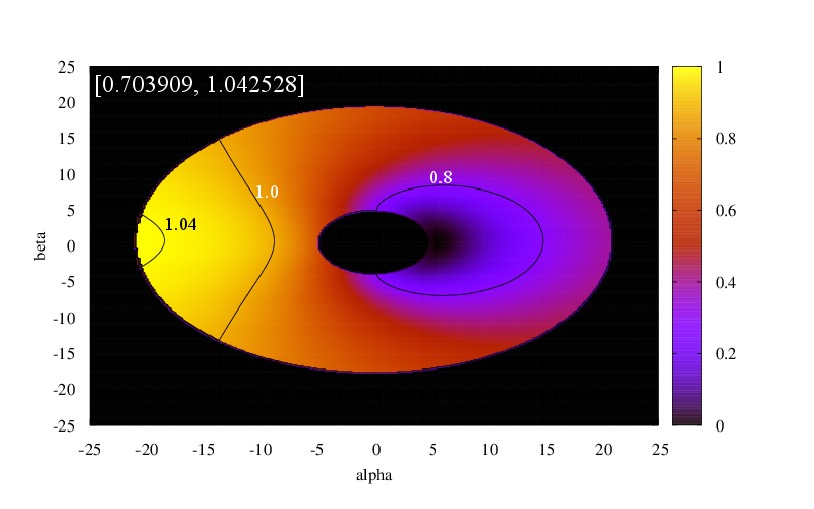}
\end{tabular}
\caption{Frequency shift map on the Keplerian disc primary image. Images are constructed for three representative values of specific charge parameter $g_{II}=(g_{NoH}+g_P)/2$ (top), $g_{III}=(g_{P}+g_S)/2$, and $1.5$ (bottom) and comparison is made between Bardeen (left) and ABG (right) spacetimes. The observer inclination is $30^\circ$. The inner edge is at $r_{ISCO}$ in the first two cases and at $r_{\Omega}$ in the case of $g=1.5$. }
\end{center}
\end{figure}

\begin{figure}[H]
	\begin{center}
		\begin{tabular}{c}
			\includegraphics[scale=0.42]{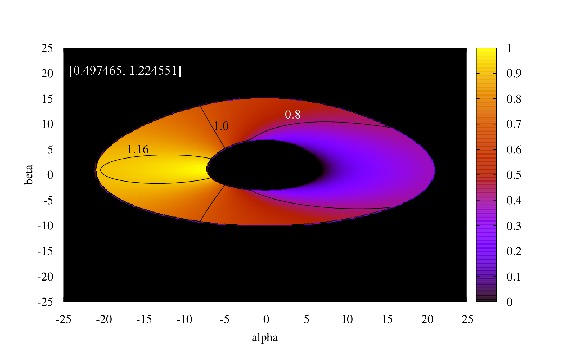}\\
			\includegraphics[scale=0.3]{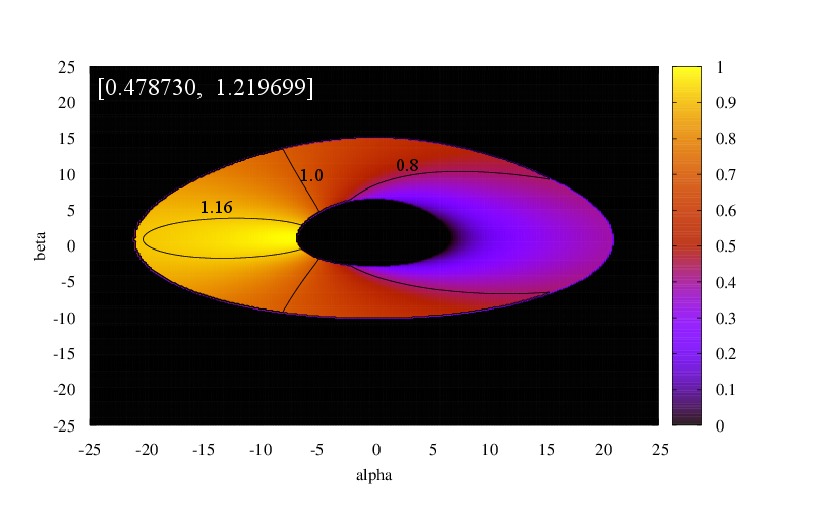}\\
			\includegraphics[scale=0.3]{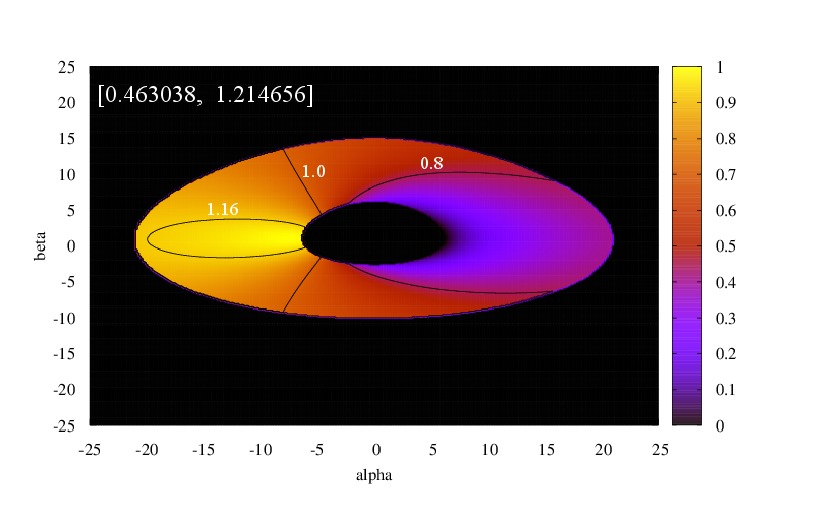}							
		\end{tabular}
		\caption{Frequency shift map on the Keplerian disc primary image. Images are constructed for Schwarzchild (top), Bardeen (middle) and ABG (bottom) black holes. The value of specific charge parameter is $g=0.5$. The observer inclination is $60^\circ$. The inner edge is at $r_{ISCO}$. }
	\end{center}
\end{figure}

\begin{figure}[H]
	\begin{center}
		\begin{tabular}{cc}
			\includegraphics[scale=0.22]{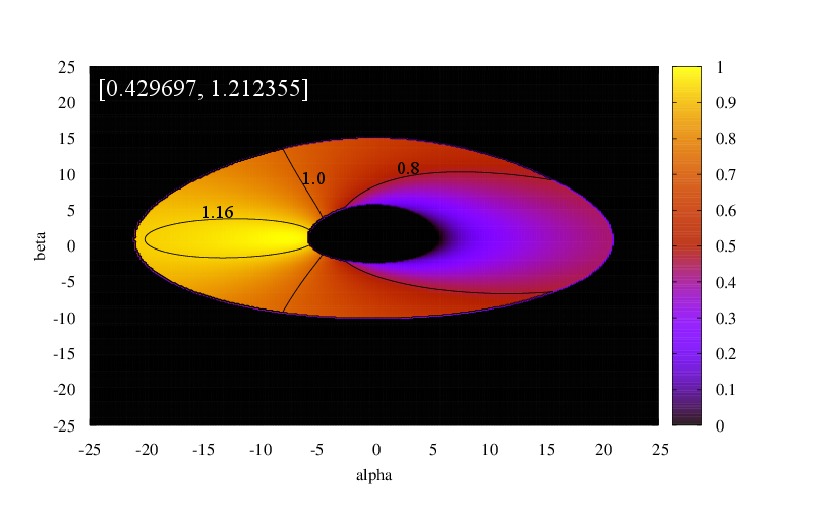}&\includegraphics[scale=0.22]{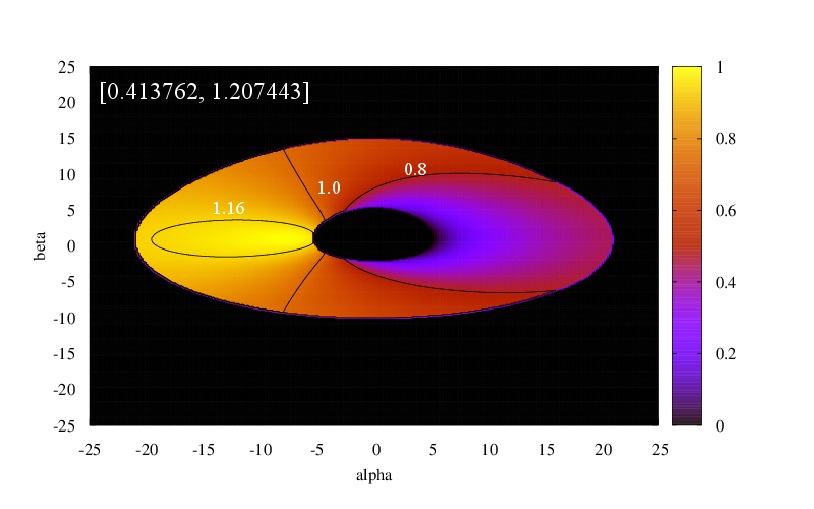}\\
			\includegraphics[scale=0.22]{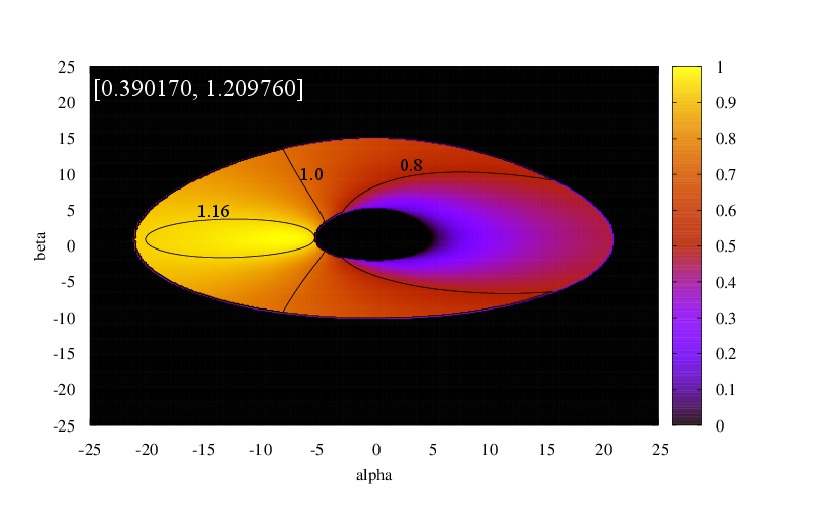}&\includegraphics[scale=0.22]{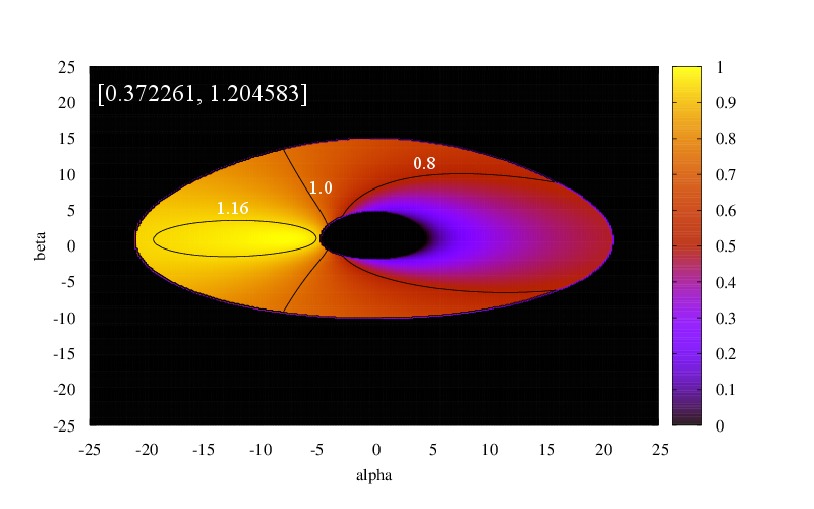}\\	
			\includegraphics[scale=0.22]{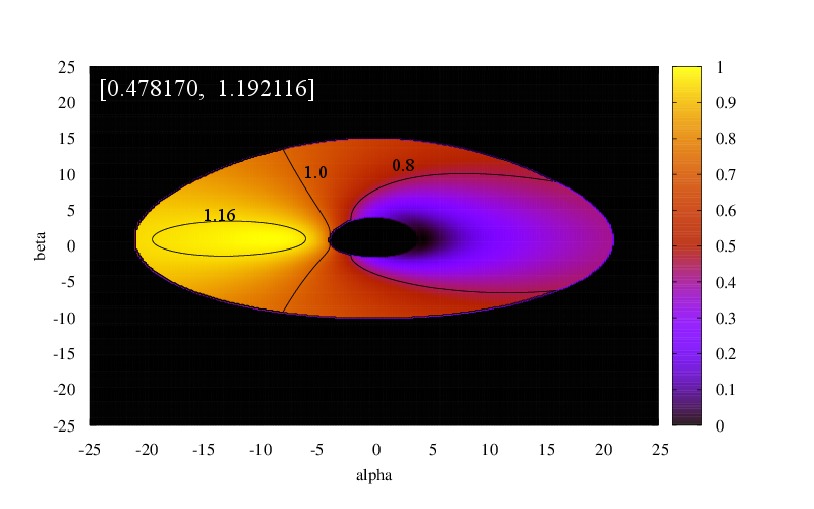}&\includegraphics[scale=0.22]{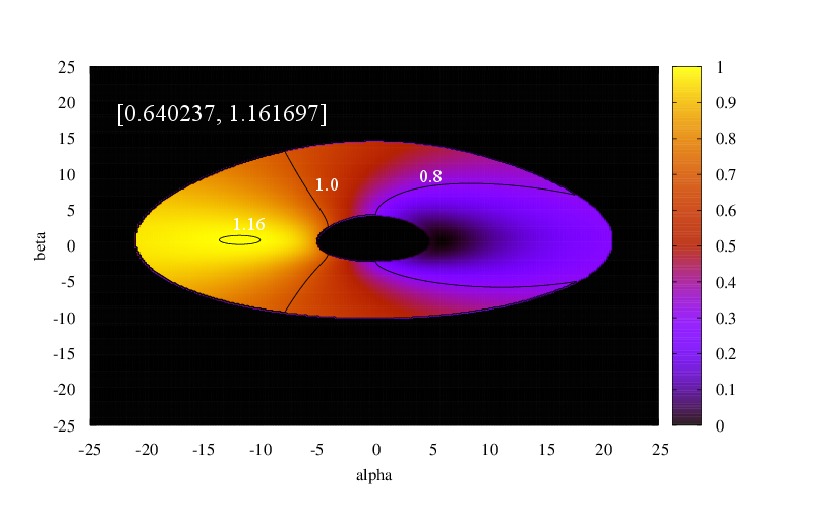}
		\end{tabular}
		\caption{Frequency shift map on the Keplerian disc primary image. Images are constructed for three representative values of specific charge parameter $g=0.5$(top), $g_{II}=(g_{NoH}+g_P)/2$, $g_{III}=(g_{P}+g_S)/2$, and $1.5$ (bottom) and comparison is made between Bardeen (left) and ABG (right) spacetimes. The observer inclination is $60^\circ$. The inner edge is at $r_{ISCO}$ in the first two cases and at $r_{\Omega}$ in the case of $g=1.5$.}
	\end{center}
\end{figure}

\begin{figure}[H]
	\begin{center}
		\begin{tabular}{c}
			\includegraphics[scale=0.42]{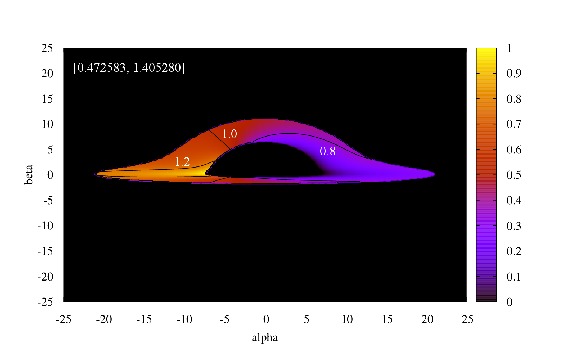}\\
			\includegraphics[scale=0.3]{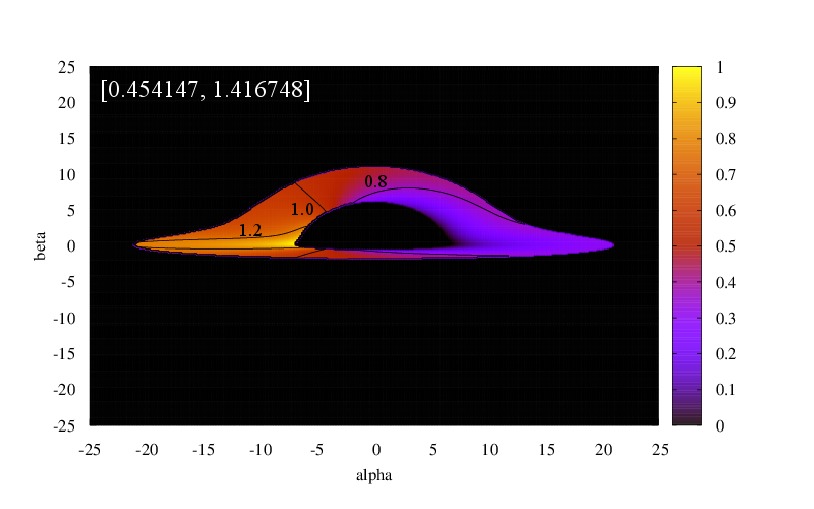}\\
			\includegraphics[scale=0.3]{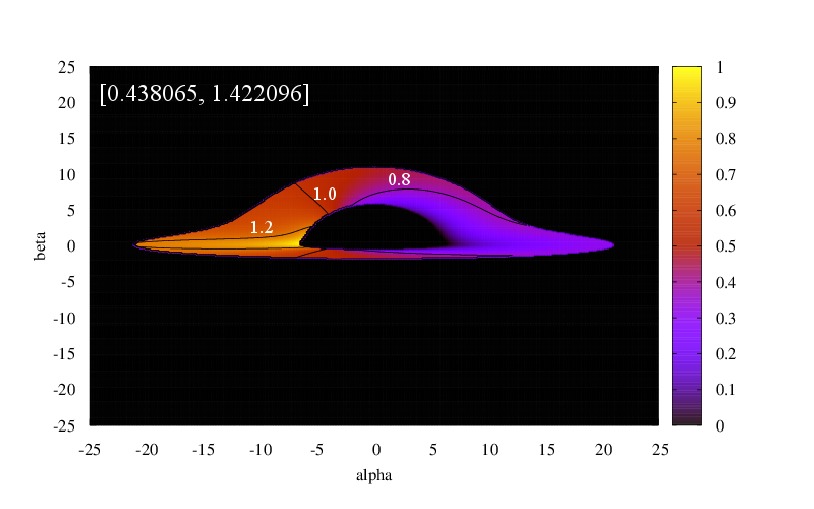}							
		\end{tabular}
		\caption{Frequency shift map on the Keplerian disc primary image. Images are constructed for Schwarzchild (top), Bardeen (middle) and ABG (bottom) black holes. The value of specific charge parameter is $g=0.5$ The observer inclination is $85^\circ$. The inner edge is at $r_{ISCO}$.}
	\end{center}
\end{figure}

\begin{figure}[H]
	\begin{center}
		\begin{tabular}{cc}
			\includegraphics[scale=0.22]{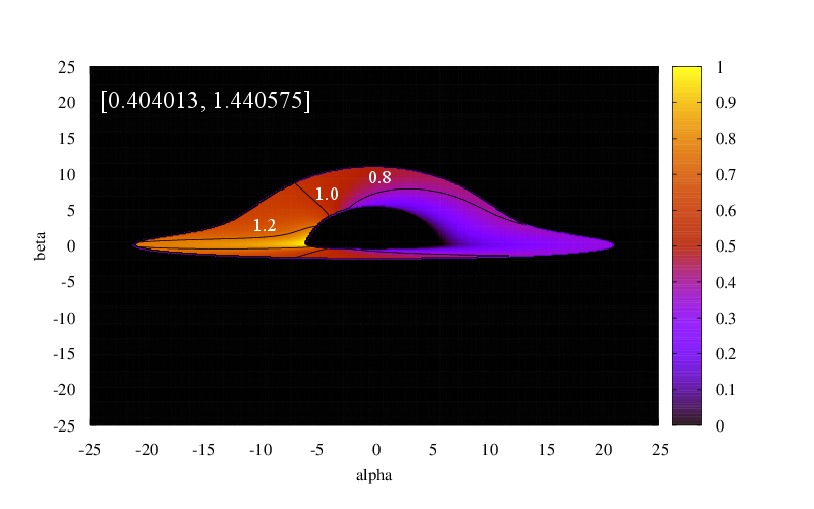}&\includegraphics[scale=0.22]{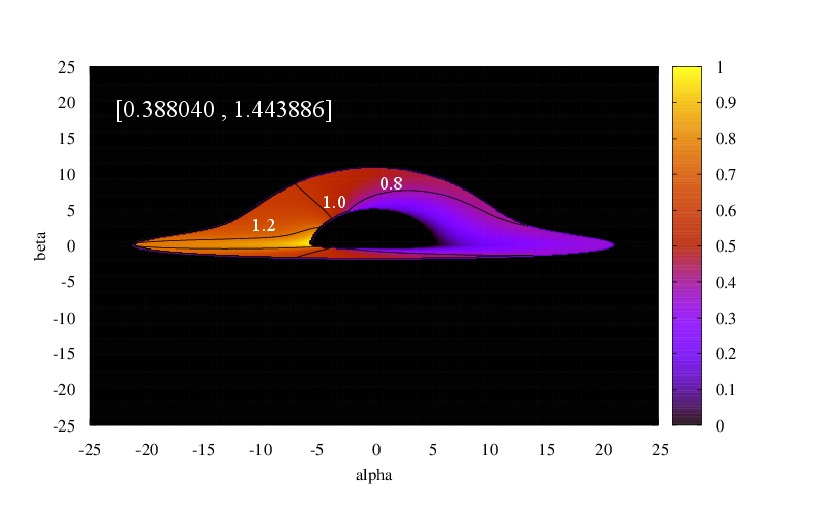}\\
			\includegraphics[scale=0.22]{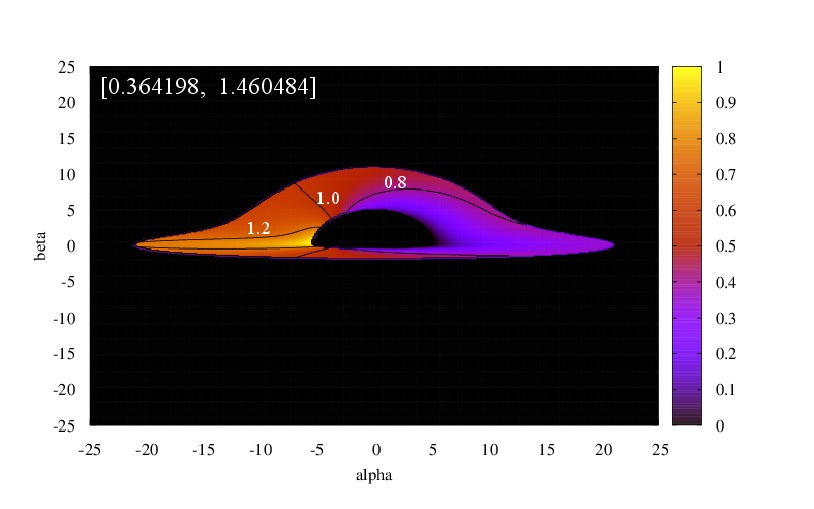}&\includegraphics[scale=0.22]{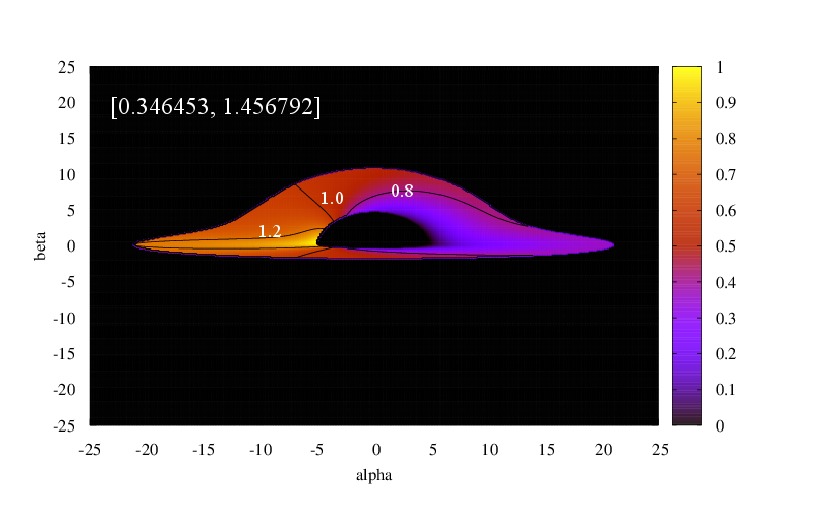}\\	
			\includegraphics[scale=0.22]{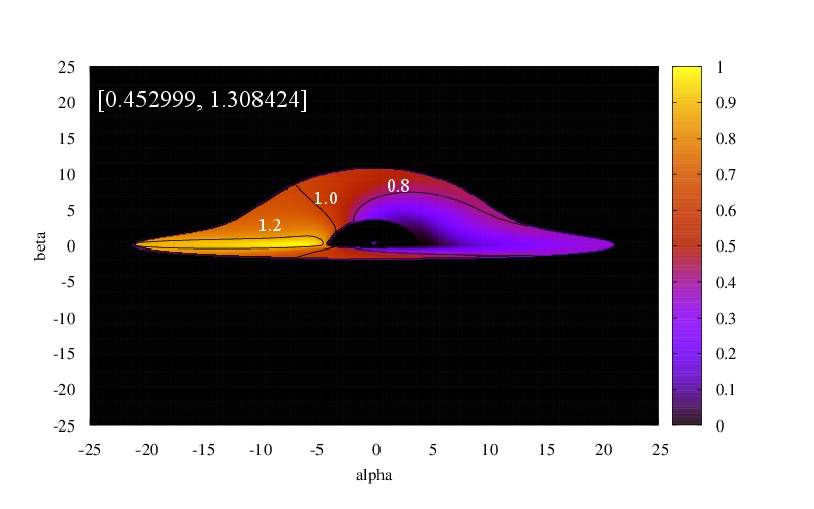}&\includegraphics[scale=0.22]{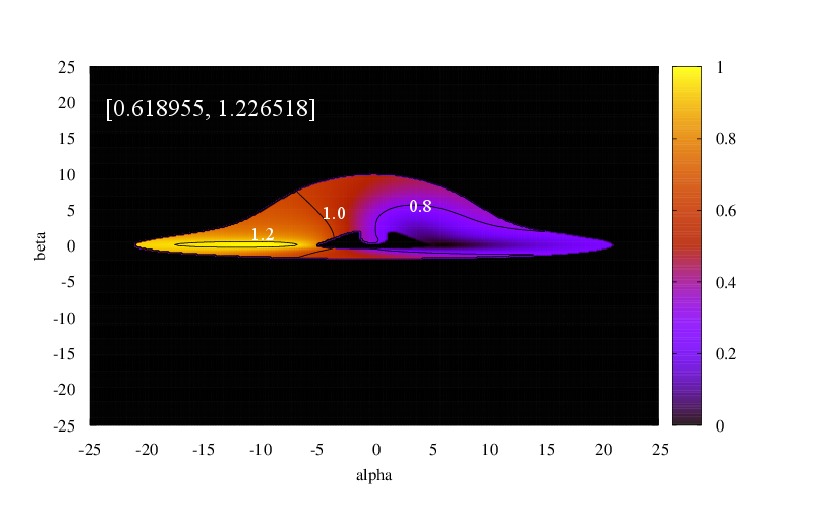}
		\end{tabular}
		\caption{Frequency shift map on the Keplerian disc primary image. Images are constructed for three representative values of specific charge parameter $g_{II}=(g_{NoH}+g_P)/2$ (top), $g_{III}=(g_{P}+g_S)/2$, and $1.5$ (bottom) and comparison is made between Bardeen (left) and ABG (right) spacetimes. The observer inclination is $85^\circ$. The inner edge is at $r_{ISCO}$ in the first two cases and at $r_{\Omega}$ in the case of $g=1.5$.}
	\end{center}
\end{figure}


\begin{figure}[H]
	\begin{center}
		\begin{tabular}{c}
			\includegraphics[scale=0.4]{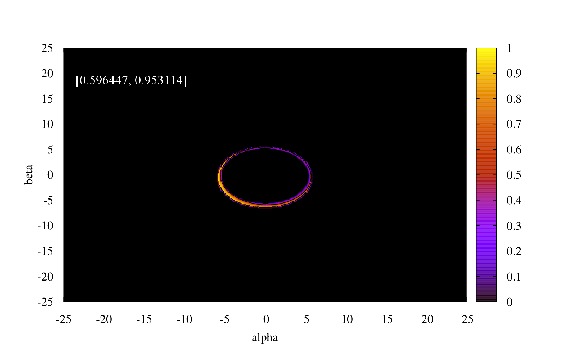}\\
			\includegraphics[scale=0.8]{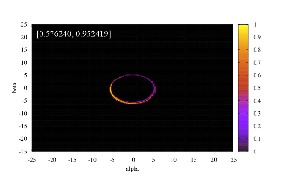}\\
			\includegraphics[scale=0.8]{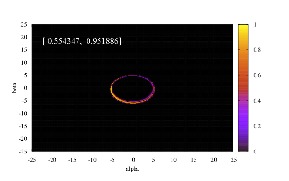}							
		\end{tabular}
		\caption{Frequency shift map on the Keplerian disc secondary image. Images are constructed for Schwarzschild (top), Bardeen (middle) and ABG (bottom) black holes. The value of specific charge parameter is $g=0.5$ The observer inclination is $30^\circ$. The inner edge is at $r_{ISCO}$. }
	\end{center}
\end{figure}

\begin{figure}[H]
	\begin{center}
		\begin{tabular}{cc}
			\includegraphics[scale=0.65]{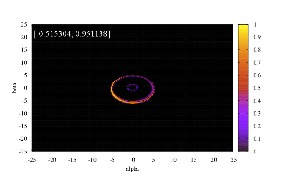}&\includegraphics[scale=0.65]{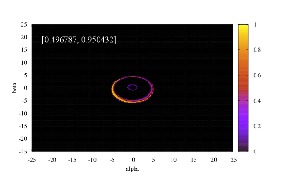}\\
			\includegraphics[scale=0.23]{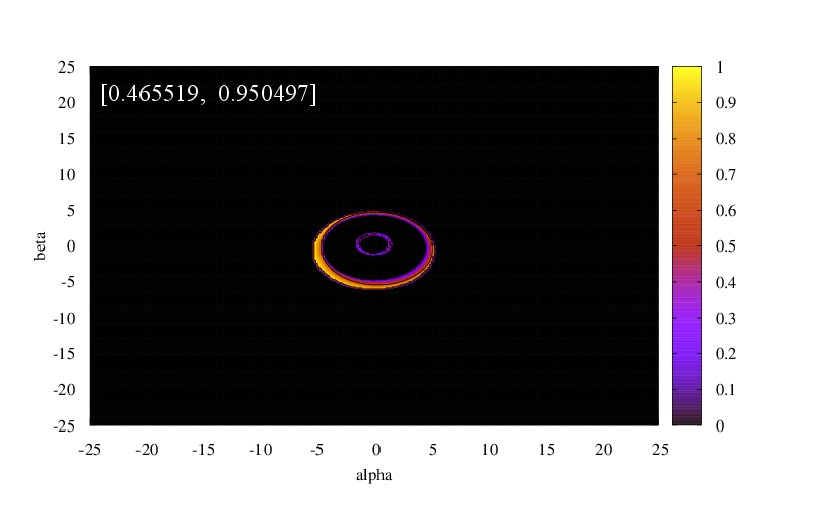}&\includegraphics[scale=0.65]{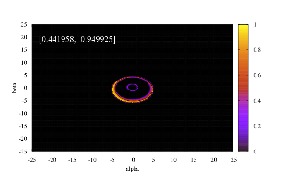}\\	
			\includegraphics[scale=0.23]{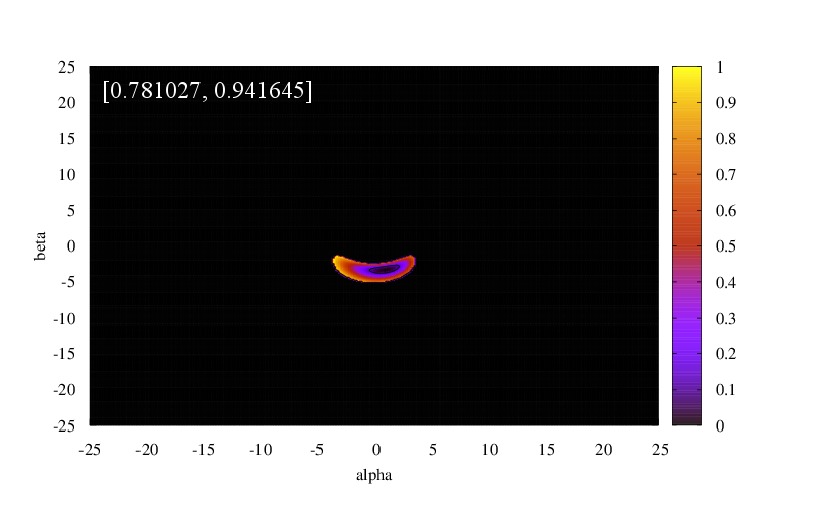}&\includegraphics[scale=0.23]{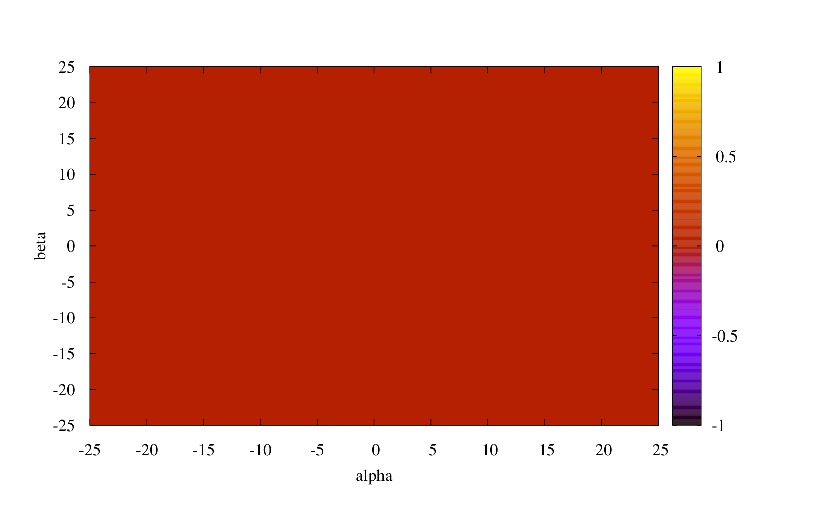}
		\end{tabular}
		\caption{Frequency shift map on the Keplerian disc secondary image. Images are constructed for three representative values of specific charge parameter  $g_{II}=(g_{NoH}+g_P)/2$ (top), $g_{III}=(g_{P}+g_S)/2$, and $1.5$ (bottom) and comparison is made between Bardeen (left) and ABG (right) spacetimes. The observer inclination is $30^\circ$. The inner edge is at $r_{ISCO}$ in the first two cases and at $r_{\Omega}$ in the case of $g=1.5$. In the $g=1.5$ Bardeen spacetime the secondary ghost image is not created, while in the $g=1.5$ ABG spacetime the secondary image can not be created at all.}
	\end{center}
\end{figure}

\begin{figure}[H]
	\begin{center}
		\begin{tabular}{c}
			\includegraphics[scale=0.4]{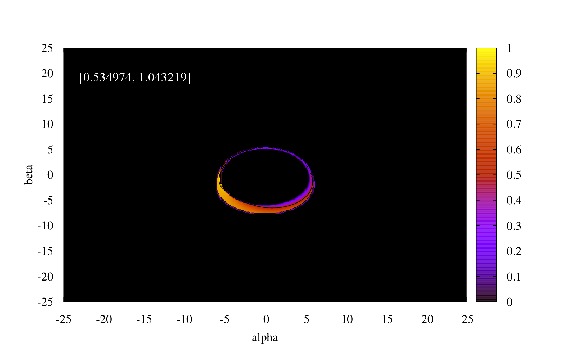}\\
			\includegraphics[scale=0.3]{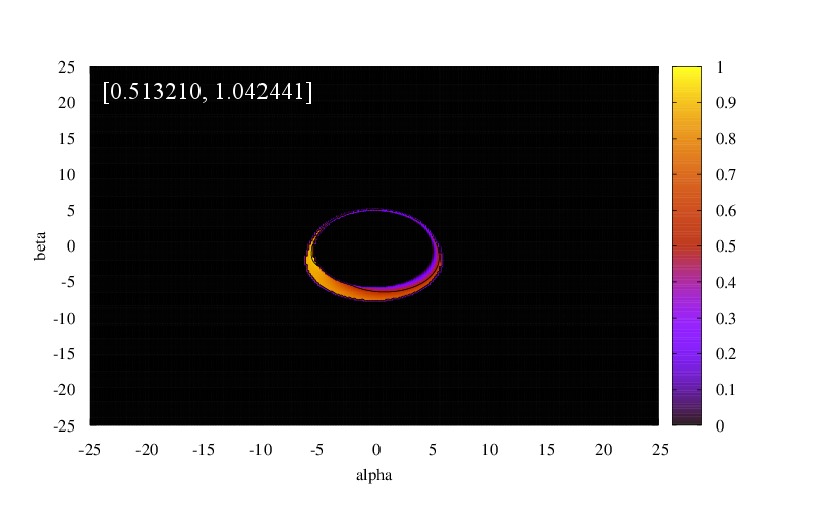}\\
			\includegraphics[scale=0.3]{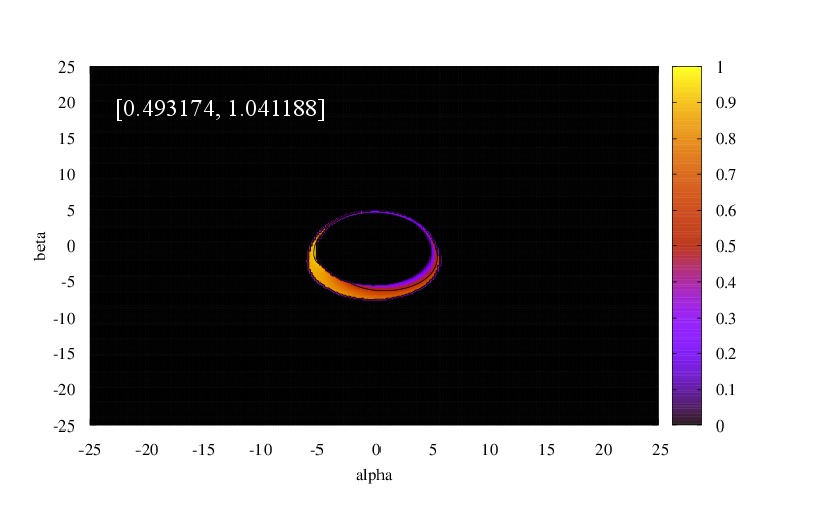}							
		\end{tabular}
		\caption{Frequency shift map on the Keplerian disc secondary image. Images are constructed for Schwarzschild (top), Bardeen (middle) and ABG (bottom) black holes. The value of specific charge parameter is $g=0.5$ The observer inclination is $60^\circ$. The inner edge is at $r_{ISCO}$.}
	\end{center}
\end{figure}

\begin{figure}[H]
	\begin{center}
		\begin{tabular}{cc}
			\includegraphics[scale=0.22]{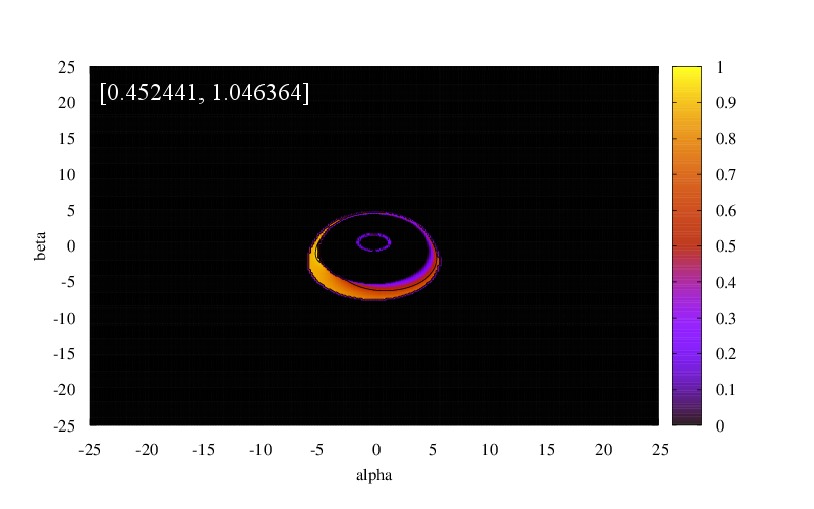}&\includegraphics[scale=0.22]{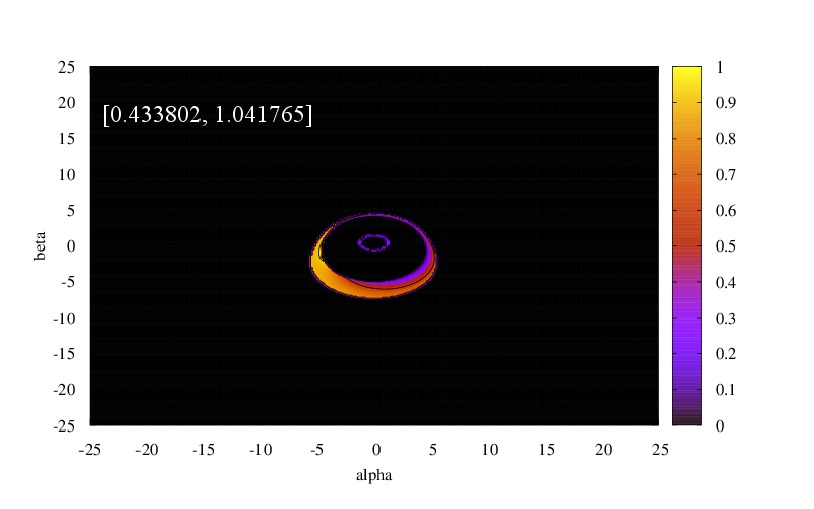}\\
			\includegraphics[scale=0.22]{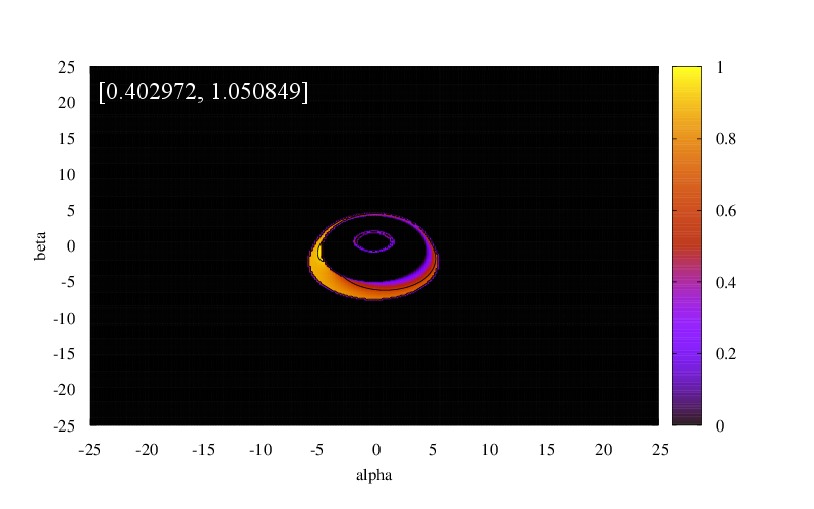}&\includegraphics[scale=0.22]{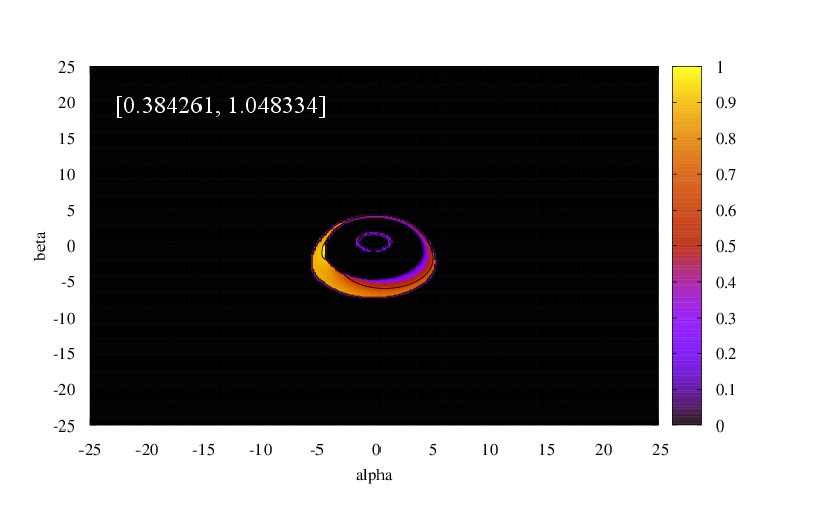}\\	
			\includegraphics[scale=0.22]{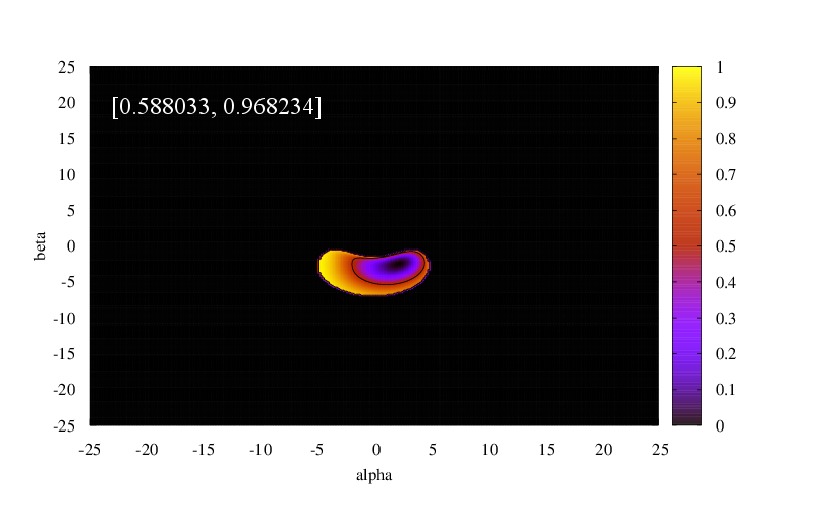}&\includegraphics[scale=0.22]{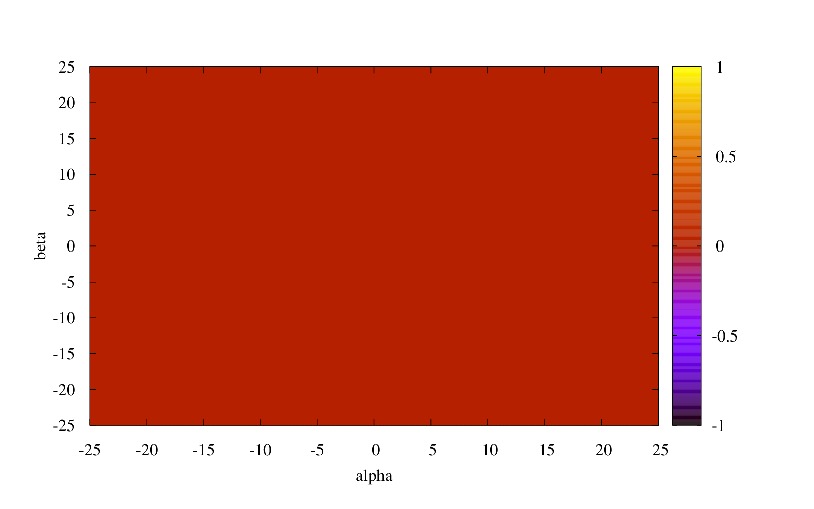}
		\end{tabular}
		\caption{Frequency shift map on the Keplerian disc secondary image. Images are constructed for three representative values of specific charge parameter  $g_{II}=(g_{NoH}+g_P)/2$ (top), $g_{III}=(g_{P}+g_S)/2$, and $1.5$ (bottom) and comparison is made between Bardeen (left) and ABG (right) spacetimes. The observer inclination is $60^\circ$. The inner edge is at $r_{ISCO}$ in the first two cases and at $r_{\Omega}$ in the case of $g=1.5$. In the $g=1.5$ Bardeen spacetime the secondary ghost image is not created, while in the $g=1.5$ ABG spacetime the secondary image can not be created at all.}
	\end{center}
\end{figure}

\begin{figure}[H]
	\begin{center}
		\begin{tabular}{c}
			\includegraphics[scale=0.45]{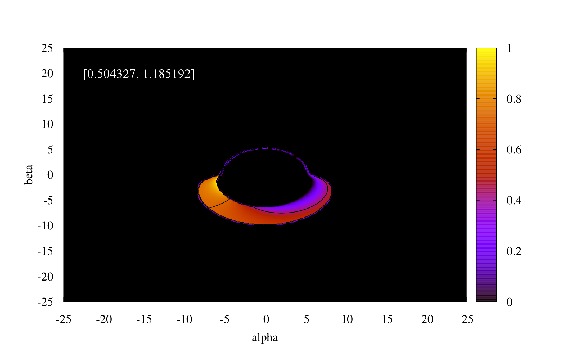}\\
			\includegraphics[scale=0.33]{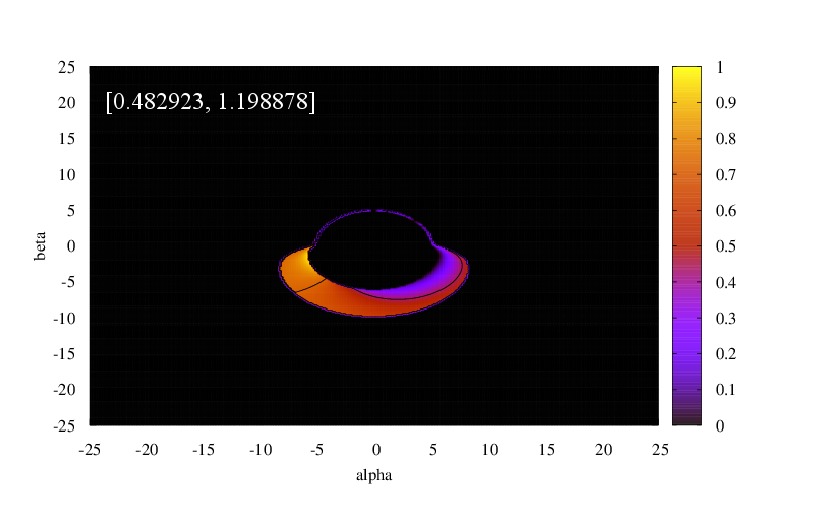}\\
			\includegraphics[scale=0.33]{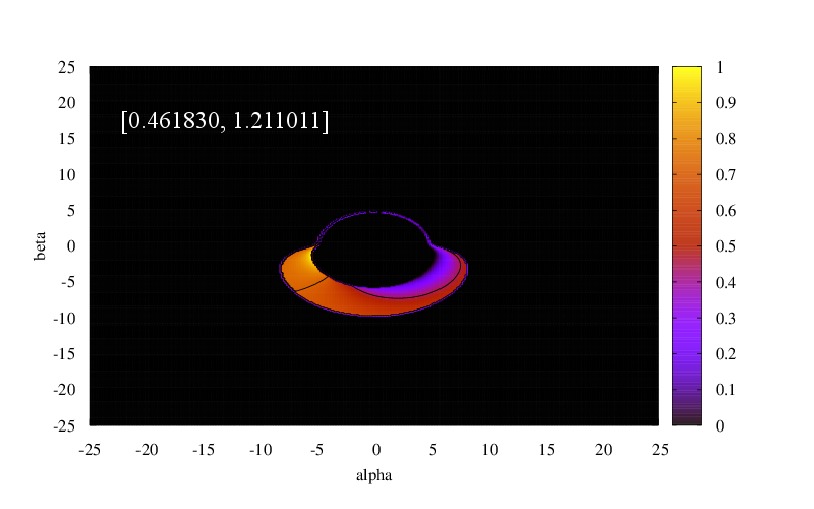}							
		\end{tabular}
		\caption{Frequency shift map on the Keplerian disc secondary image. Images are constructed for Schwarzschild (top), Bardeen (middle) and ABG (bottom) black holes. The value of specific charge parameter is $g=0.5$ The observer inclination is $85^\circ$. The inner edge is at $r_{ISCO}$. }
	\end{center}
\end{figure}

\begin{figure}[H]
	\begin{center}
		\begin{tabular}{cc}
			\includegraphics[scale=0.22]{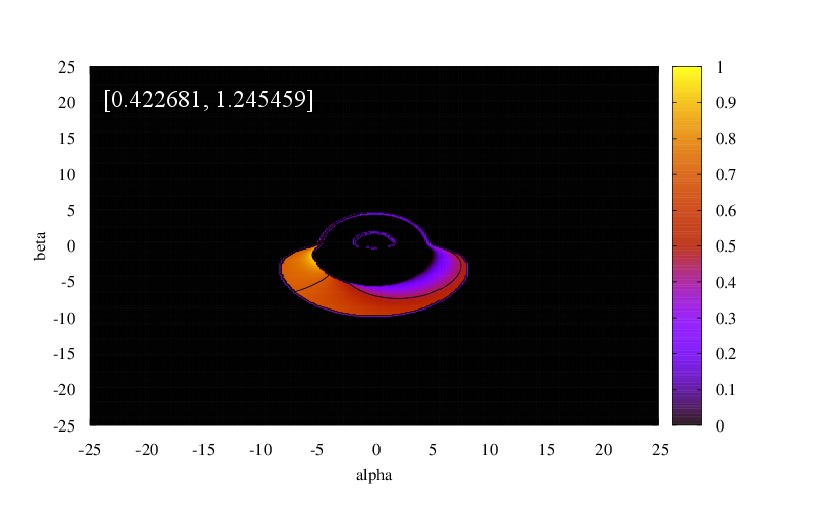}&\includegraphics[scale=0.22]{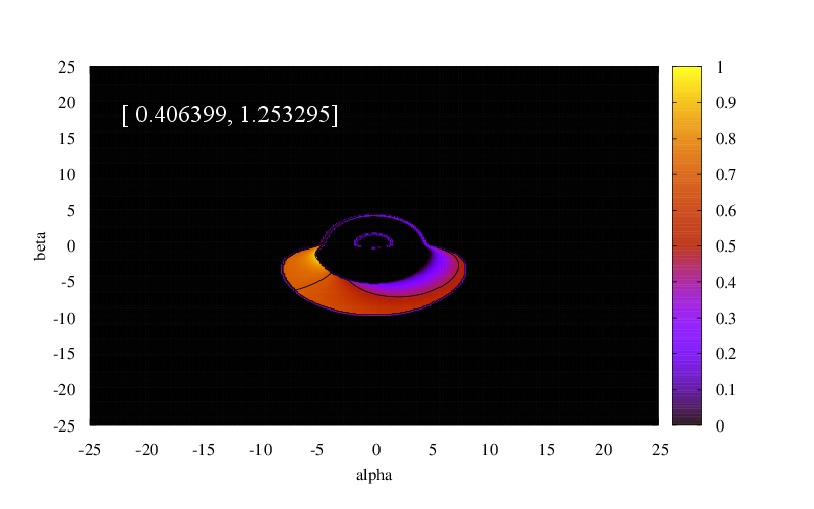}\\
			\includegraphics[scale=0.22]{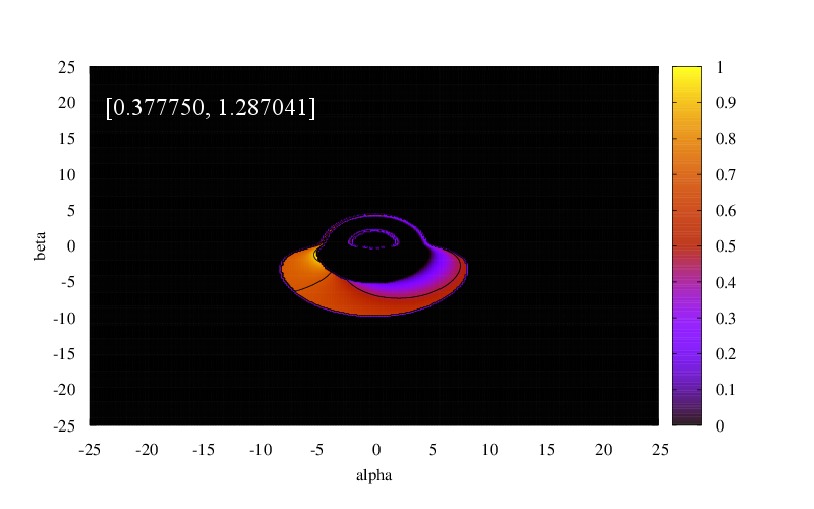}&\includegraphics[scale=0.22]{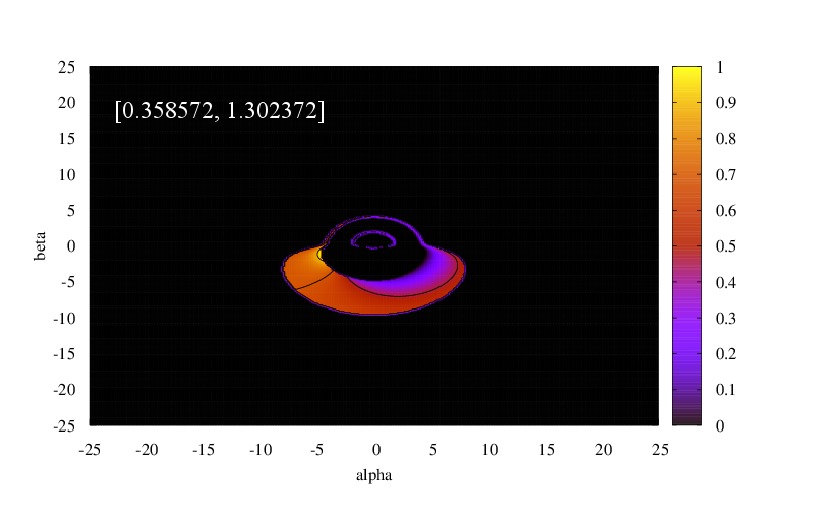}\\	
			\includegraphics[scale=0.22]{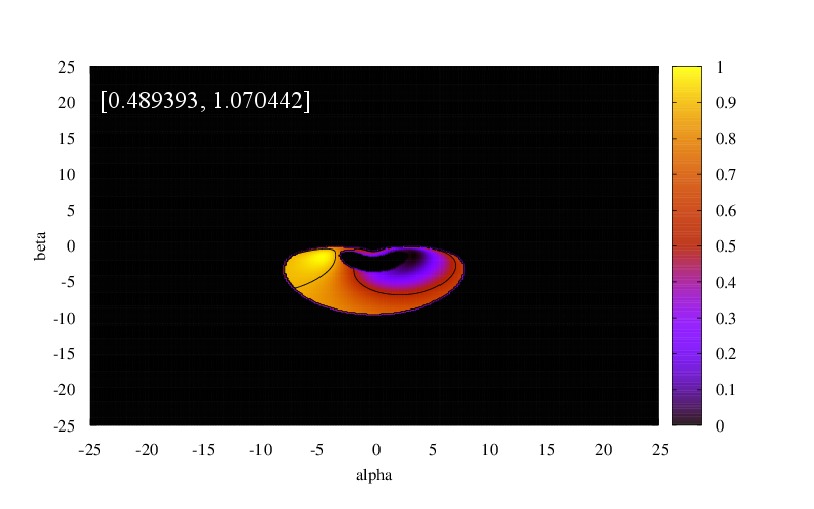}&\includegraphics[scale=0.22]{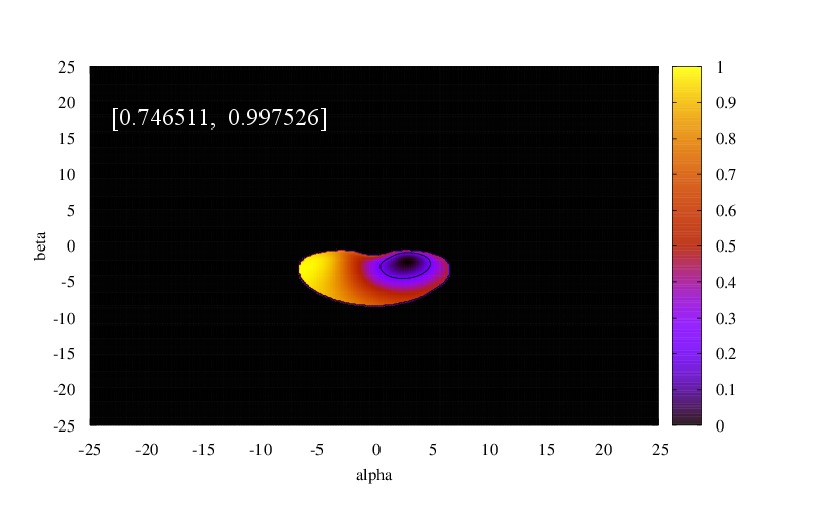}
		\end{tabular}
		\caption{Frequency shift map on the Keplerian disc secondary image. Images are constructed for three representative values of specific charge parameter  $g_{II}=(g_{NoH}+g_P)/2$ (top), $g_{III}=(g_{P}+g_S)/2$, and $1.5$ (bottom) and comparison is made between Bardeen (left) and ABG (right) spacetimes. The observer inclination is $85^\circ$. The inner edge is at $r_{ISCO}$ in the first two cases and at $r_{\Omega}$ in the case of $g=1.5$. For $g=1.5$, in both Bardeen and ABG spacetimes the secondary ghost images are not created.}
	\end{center}
\end{figure}

\begin{figure}[H]
	\begin{center}
	\begin{tabular}{cc}
		\includegraphics[scale=0.22]{image_1_4_bardeen_I}&\includegraphics[scale=0.31]{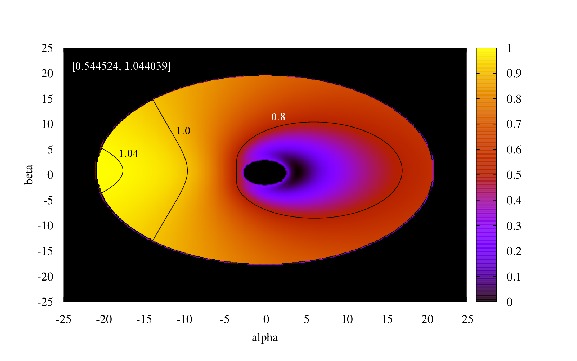}\\
		\includegraphics[scale=0.22]{image_2_4_bardeen_I}&\includegraphics[scale=0.31]{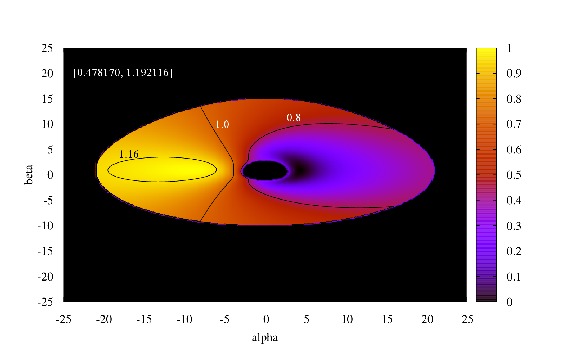}\\
		\includegraphics[scale=0.22]{image_3_4_bardeen_I}&\includegraphics[scale=0.31]{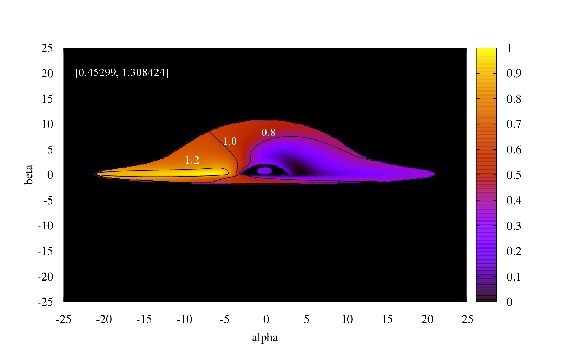}
	\end{tabular}
	\caption{Frequency shift map of the Keplerian disc primary image constructed in the Bardeen spacetime determined by the magnetic charge parameter $g=1.5$, spanning from $r_{in}=r_{\Omega max}$ (left), $r_{stat}$ (right) respectively to $r_{out}=20 M$ . The inclination of observer is $\theta_o=30^\circ$ (top), $60^\circ$ (middle), and $85^\circ$ (bottom). Notice occurence of the ghost image in the disc with $r_{in} = r_{stat}$.}
	\end{center}
\end{figure}

\begin{figure}[H]
	\begin{center}
		\begin{tabular}{cc}
			\includegraphics[scale=0.22]{image_1_4_bardeen_II}&\includegraphics[scale=0.31]{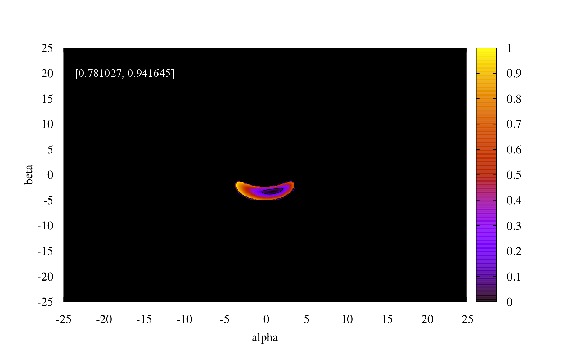}\\
			\includegraphics[scale=0.22]{image_2_4_bardeen_II}&\includegraphics[scale=0.31]{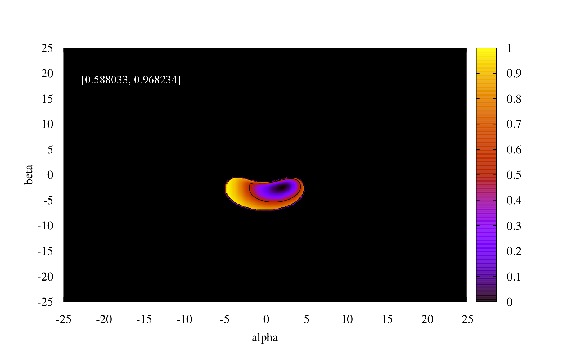}\\
			\includegraphics[scale=0.22]{image_3_4_bardeen_II}&\includegraphics[scale=0.31]{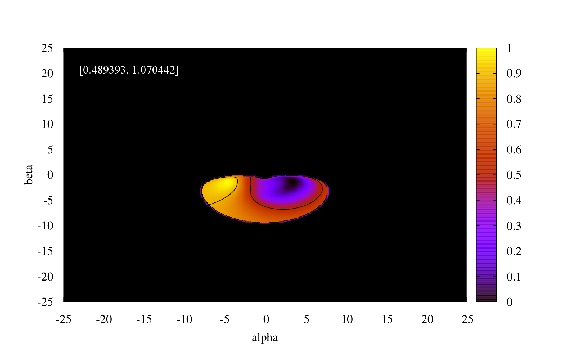}
		\end{tabular}
		\caption{Frequency shift map of the Keplerian disc secondary image constructed  in Bardeen spacetime determined by magnetic charge parameter $g=1.5$ spanning from $r_{in}=r_{\Omega max}$ (left), $r_{stat}$ (right) respectively to $r_{out}=20 M$ . The inclination of observer is $\theta_o=30^\circ$ (top), $60^\circ$ (middle), and $85^\circ$ (bottom).}
	\end{center}
\end{figure}

We see that the images created in the Bardeen and ABG black hole spacetimes are of the same character as those created by the Schwarzschild black hole. Small differences are reflected by the range of the frequency shift - they have the same character for both direct and indirect images and considered inclination angles. For small ($30^\circ$) and intermediate ($60^\circ$) inclination the frequency range shifts to smaller values at both ends of the range, as compared to the Schwarzschild range; the shift is larger for the ABG black holes than for the Bardeen black holes. For large inclination angles ($85^\circ$), the frequency shift decreases at the red end of the range and increases at the blue end, being larger for the ABG black holes. Generally, the extension of the frequency range is larger for direct images as compared to the indirect images.

We further observe that the direct images have the same character for the Bardeen and ABG spacetimes of Class II-III in both the image shape and the frequency shift distribution. Clear differences arise in the case of the Class IV spacetimes, with inner edge of the Keplerian disc at $r_{in}=r_{\Omega}$. For a given spacetime parameter $g=1.5$, the ghost images are more visible in the ABG spacetime, while the range of the frequency shift across the direct image is narrower in comparison to those in the Bardeen spacetime, as demonstrated especially for the large inclination angle (Figure 7) -- we can see even the ghost image in the ABG spacetime, while it does not occur in the Bardeen spacetime. Note that the highest blueshift occurs for the direct images created in the black hole spacetimes for the small and mediate inclination angles, while for the large inclination the largest blueshift occurs for the direct images generated in the Bardeen Class III spacetime. 

For the indirect images only minor shift of their properties occurs again in the Bardeen and ABG spacetimes of Class II-III. For the Class IV ABG spacetime with $g=1.5$, the indirect images do not exist for small and mediate inclination angles (Figures 9 and 11) as their gravitational field is too weak to cause deflection of light strong enough to create the indirect images under small and mediate inclination angles. The ghost secondary images do not exist in both the Bardeen and ABG Class IV spacetimes for the large inclination angle (Figure 13). 

For values of $g>1.5$, properties of the images in the Bardeen spacetimes can be similar to those demonstrated in the $g=1.5$ ABG spacetime, as shown in \cite{Sche-Stu:2015:JCAP:} where also the higher-order images in the Bardeen Class II spacetimes have been studied. We shall not repeat here discussion of the higher-order images that is qualitatively the same for the Bardeen and ABG spacetimes. For the highest blueshift in the indirect images we observe the same qualitative properties as those that are demonstrated in the direct images. Generally, the highest blueshift is lower in the indirect images than in the direct images. 

The possible influence of the innermost parts of Keplerian discs in the Bardeen spacetimes located at $r_{stat}<r<r_{\Omega}$ is reflected for direct (Figure 14) and indirect (Figure 15) images that demonstrate no effect on frequency shift range, but for large inclination angle ($85^\circ$) the occurrence of the direct ghost image is demonstrated in Figure 14.

\section{Profiles of spectral lines}

In constructing the profiled spectral lines we assume the radiation that originates in the innermost regions of Keplerian disc governed by the strong gravity of the regular black hole and no-horizon spacetimes. The disc is composed of point sources orbiting on circular geodetical orbits and radiating locally isotropicaly and monochromatically, i.e., at the frequency given by a considered spectral line, usually the Fe spectral line giving X-ray radiation. The spectral line is then profiled by the gravitational lensing and by the gravitational frequency shift combined with the Doppler frequency shift due to the orbital motion that are related to a fixed distant observer.  

The profiles of spectral lines are constructed for the radiation coming from the disc region between the inner edge at $r_{in}$ and some appropriatelly chosen outer edge at $r_{out}$. Each emitted photon suffers from gravitational and Doppler frequency shift which takes the general form
\beq
	\mathcal{G}=\frac{\sqrt{f-r^2\Omega^2}}{1-l \Omega},
\eeq 
where the photon impact parameter $l=-p_\phi/p_t=L/E$ and $\Omega$ is the Keplerian angular frequency. 

The specific flux at the detector $F_{\nu0}$ is constructed by binning the photons (pixels) contributing to specific flux $F$ at observed frequency $\nu_0$. Let i-th pixel on the detector subtends the solid angle $\Delta\Pi_i$. Then the corresponding flux $\Delta F_i(\nu_0)$ reads
\beq
	\Delta F_i(\nu_o)=I_{o}(\nu_o)\Delta\Omega_i=\mathcal{G}_i^3 I_e(\nu_o/\mathcal{G}_i) \Delta\Pi_i,
\eeq
where the specific intensity of naturally (thermally) broadened line with the power law emissivity model is given by 
\beq
	I_e=\epsilon_o r^{-p} \exp[-\gamma(\nu_o/\mathcal{G} -\nu_0)^2].
\eeq
In our simulations the dimensionless parameter $\gamma=10^3$ and the emissivity law index $p=2$. 
 
The solid angle is given by the coordinates $\alpha$ and $\beta$ on the observer plane due to the relation $d\Pi =  d\alpha d\beta/{D^{2}_{\rm o}}$, where $D_{\rm o}$ denotes the distance to the source. The coordinates $\alpha$ and $\beta$ can be then expressed in terms of the radius $r_{\rm e}$ of the source orbit and the related redshift factor $\mathcal{G}=\nu_{\rm o}/\nu_{\rm e}$. The Jacobian of the transformation $(\alpha,\beta) \rightarrow (r_{\rm e},g)$ implies \cite{Sche-Stu:2009:GenRelGrav:,Sche-Stu:2013:JCAP:}
\begin{equation}
d\Pi=\frac{q}{D_{\rm o}^2\sin\theta_{\rm o}\sqrt{q-\lambda^2\cot^2\theta_{\rm o}}}\left|\frac{\partial r_{\rm e}}{\partial\lambda}\frac{\partial \mathcal{G}}{\partial q}-\frac{\partial r_{\rm e}}{\partial q}\frac{\partial \mathcal{G}}{\partial \lambda}\right|^{-1}\!\!\!\!\!{d} \mathcal{G}{d} r_{\rm e} \rightarrow \Delta\Pi_i.
\end{equation}
The parameter $q$ represents the total photon impact parameter related to the plane of motion of the photon, while $\lambda$ represents the axial impact parameter related to the plane of the Keplerian disc. To obtain the specific flux at a particular frequency $\nu_o$, all contributions given by $\Delta F_i(\nu_o)$ are summed 
\beq
	F(\nu_o)=\sum\limits_{i=0}^n \Delta F(\nu_o)_i.
\eeq  

\section{Constructed profiled lines}

We construct the profiled lines assuming in the standard way the extension of the radiating Keplerian discs to be restricted between the innermost stable circular orbit at $r_{in}=r_{ISCO}$ and $r_{out}=20m$ for the black hole spacetimes. In the case of the no-horizon spacetimes with $(g/m)_{NoH/B} < g/m < (g/m)_{S/B}$ ($(g/m)_{NoH/ABG} < g/m < (g/m)_{S/ABG}$), i.e., with doubled region of the stable circular orbits, we do not consider the inner region of the stable orbits and we put $r_{in}=r_{ISCO}$. In the case of the no-horizon spacetimes with $g/m > (g/m)_{S/B}$ ($g/m > (g/m)_{S/ABG}$), we do not consider the region of $r < r_{\Omega}$ and we put $r_{in}= r_{\Omega}$. However, in a special case we demonstrate the role of the contribution from the radiating region $r_{stat}<r<r_{\Omega}$ to the profiled spectral line.  On the other hand, we include the contribution of the ghost images \cite{Sche-Stu:2015:JCAP:} to the profiled spectral lines. In all the no-horizon spacetimes, we keep the outer edge of the radiating disc at $r_{out}=20m$. We compare our results to the profiled line generated by the Keplerian disc with $r_{in}=r_{ISCO}$ and $r_{out}=20m$ orbiting a Schwarzschild black hole.\footnote{We shall not consider here for simplicity the effect of light emitted within the ISCO that can influence spectra and image of accreation discs as shown recently in magnetohydrodynamics calculations. \cite{Nob-Kro-Schn-Haw:2011:ApJ:,Zhu-etal:2012:MNRAS:,Schn-Kro-Nob:2013:ApJ:}}.   

\subsection{Black hole spacetimes}
The modelled profiles of spectral lines generated by the Keplerian discs in vicinity of the Bardeen and ABG black holes are constructed for the spacetimes with the same charge parameter $g/m=0.5$ and the results are presented in Figure 16 for the typical inclination angles of the Keplerian discs to the distant observer $\theta_{o} = 30^{\circ}$, $60^{\circ}$, $85^{\circ}$. The same angles will be used for construction of the profiled lines in all the considered no-horizon spacetimes. 

\begin{figure}[H]
	\begin{center}
		\begin{tabular}{ccc}
			\includegraphics[scale=0.3]{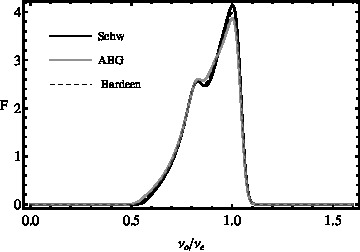}&
			\includegraphics[scale=0.3]{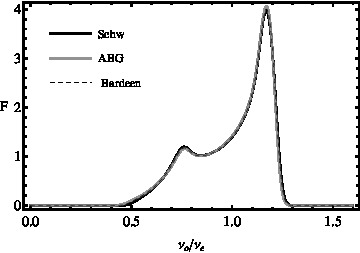}&
			\includegraphics[scale=0.3]{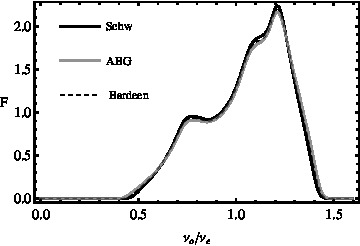}
		\end{tabular}
		\caption{The profile of spectral line generated in Keplerian disc orbiting in the vicinity of black holes. The parameter $g=0.5$ is chosen for both the Bardeen and ABG spacetimes. The observer inclination is $\theta_o=30^\circ$ (top), $60^\circ$ (middle), and $85^\circ$ (bottom). \label{pl_bh}}
	\end{center}
\end{figure}

We can see that for both the Bardeen and ABG spacetimes the profiled lines follow very closely the line corresponding to the Schwarzschild black hole case. It is quite interesting that the coincidence is strongest for the mediate inclination angle ($\theta_{o} = 60^{\circ}$). Small differences arise for the small ($\theta_{o} = 30^{\circ}$) and large ($\theta_{o} = 85^{\circ}$) inclination angles. For both small and large inclination angles the tendency in both the Bardeen and ABG spacetimes is to make the profile flatter in comparison to the Schwarzschild profile; for increasing charge parameter the peak height is slightly suppressed, while the frequency range slightly increases. The modifications in the ABG black hole spacetime are slightly stronger than those in the Bardeen spacetime. However, these modifications are very small and their detectability is probably out of abilities of recent observational technique. 



\subsection{No-horizon spacetimes}
We have constructed the profiled lines for the three Classes (II-IV) of the no-horizon regular Bardeen and ABG spacetimes. As in the black hole case, we compare the profiled lines for a specific value of the charge parameter of the Bardeen and ABG spacetimes. For the no-horizon spacetimes of the Class III and IV we then demonstrate the role of increasing charge parameter $g/m$. 

\subsubsection{Spacetimes with trapped null geodesics}
In the case of the Bardeen (ABG) no-horizon spacetimes with $g<(g/m)_{P/B}$ ($g<(g/m)_{P/ABG}$), the profiles of spectral lines take the form demonstrated in Figure 17. 

\begin{figure}[H]
	\begin{center}
		\begin{tabular}{ccc}
			\includegraphics[scale=0.3]{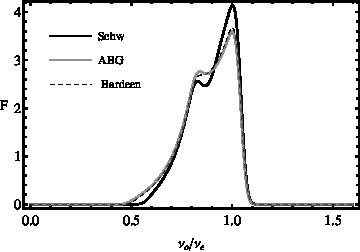}&
			\includegraphics[scale=0.3]{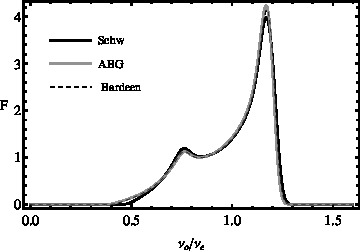}&
			\includegraphics[scale=0.3]{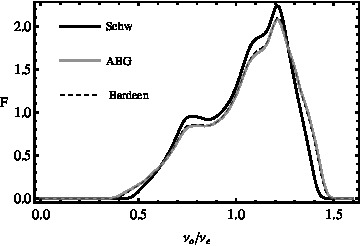}
		\end{tabular}
		\caption{The profile of spectral line generated in the Keplerian disc formed in the Schwarzschild black hole and Bardeen and ABG no-horizon spacetimes of Class II. The parameter $g=\frac{1}{2}(g_{NoH}+g_{p})$ in both Bardeen and ABG spacetimes. The observer inclination is $\theta_o=30^\circ$ (top), $60^\circ$ (middle), and $85^\circ$ (bottom). \label{pl_noh-p}}
	\end{center}
\end{figure}

For the small ($\theta_{o} = 30^{\circ}$) and large ($\theta_{o} = 85^{\circ}$) inclination angles the modifications of the Bardeen and ABG lines relative to the Schwarzschild line have the same character as in the black hole case, being more profound. The profile height is always reduced, the frequency range is increased at the red-end and slightly decreased at the blue-end for the small inclination, while it is increased at both ends of the frequency range for large inclination. For mediate inclination ($\theta_{o} = 60^{\circ}$), the height 
of the profiled line increased for the Bardeen spacetime, but decreased for the ABG spacetime, as related to the height of the profiled line in the Schwarzschild spacetime. The frequency range increases at the red-end for both the spacetime, while at the blue-end it decreases (increases) for the Bardeen (ABG) spacetime. Very precise measurements only could enable distinction of the profiled lines in the Bardeen and ABG spacetimes. 

\subsubsection{Spacetimes with unstable circular geodesics}
In the case of the Bardeen and ABG no-horizon spacetimes with $g<(g/m)_{S/B}$ ($g<(g/m)_{S/ABG}$), the profiled spectral lines are compared in Figure 18. Recall that now the inner edge of the Keplerian discs corresponds to the outer marginally stable orbit again. The influence of increasing spacetime charge parameter is demonstrated in Figure 19. 

\begin{figure}[H]
	\begin{center}
		\begin{tabular}{ccc}
			\includegraphics[scale=0.3]{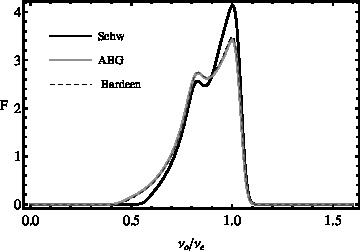}&
			\includegraphics[scale=0.3]{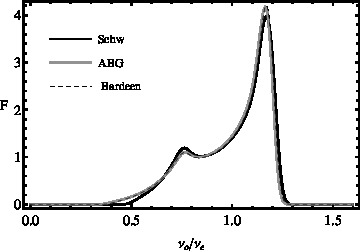}&
			\includegraphics[scale=0.3]{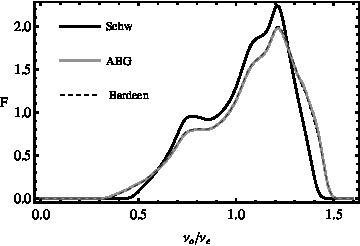}
		\end{tabular}
		\caption{The profile of spectral line generated in the Keplerian disc orbiting in the  Schwarzschild black hole and Bardeen and ABG no-horizon spacetimes of Class III. The parameter $g=\frac{1}{2}(g_{p}+g_{s})$ in both the Bardeen and ABG spacetimes. The observer inclination is $\theta_o=30^\circ$ (top), $60^\circ$ (middle), and $85^\circ$ (bottom).\label{pl_p-s} }
	\end{center}
\end{figure}

\begin{figure}[H]
	\begin{center}
		\begin{tabular}{ccc}
			\includegraphics[scale=0.4]{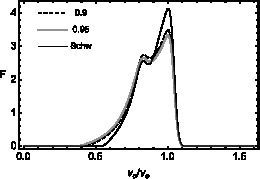}&\includegraphics[scale=0.4]{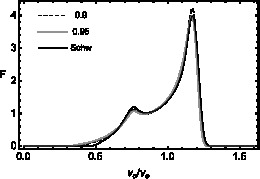}&\includegraphics[scale=0.4]{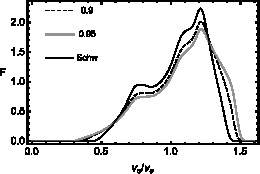}\\
			\includegraphics[scale=0.4]{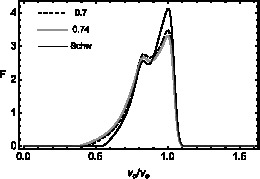}&\includegraphics[scale=0.4]{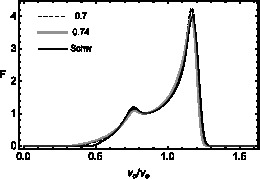}&\includegraphics[scale=0.4]{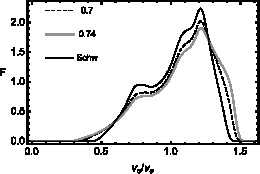}
		\end{tabular}
		\caption{Profiles of spectral lines of radiation generated in the vicinity of Bardeen (top) and ABG (bottom) no-horizon Class III spacetimes for two representative values of the charge parameter $g=0.9$(dashed, top) and $0.95$(dotted, top), and for $g=0.7$(dashed, bottom) and $0.74$(dotted, bottom). The inclination of observer is $\theta_o=30^\circ$(left), $60^\circ$(middle), and $85^\circ$(right). The black solid curves correspond to the profiled line created in vicinity of the Schwarzchild black hole.}
	\end{center}
\end{figure}

Now we can conclude that the relative behavior of the Bardeen and ABG profiled lines, and their relation to the Schwarzschild line, are qualitatively the same as in the case of the Class II no-horizon spacetimes, but the differences are slightly magnified in relation to those occurring in the Class II spacetimes and could be observed easily. 

We also can see that for the ABG spacetimes increasing of the specific charge parameter makes the profiled lines flatter for all the inclination angles, the height is reduced, while the frequency range expands at both end of the range. In the Bardeen spacetimes, the same statement holds for the small and large inclination angles, while for the mediate inclination ($\theta_{o} = 60^{\circ}$) the height increases, while the frequency range decreases at the blue-end with $g/m$ increasing. In the case of the no-horizon spacetimes of the Class III the signatures of the spacetimes in the profiled lines could be in principle measurable, but they are of quatitative character only. 

\subsubsection{Spacetimes allowing only stable circular orbits}
In the case of the Bardeen and ABG Class IV spacetimes, with $g>(g/m)_{S/B}$ ($g>(g/m)_{S/ABG}$), the inner edge of the standard (MRI governed) Keplerian disc has to be located at $r_{\Omega}$. First,  we shall not include the remaining part of the disc, extending down to the static radius. The Bardeen and ABG profiles of lines are compared in Figure 20, while influence of increasing spacetime charge parameter is demonstrated in Figure 21. 

\begin{figure}[H]
	\begin{center}
		\begin{tabular}{ccc}
			\includegraphics[scale=0.3]{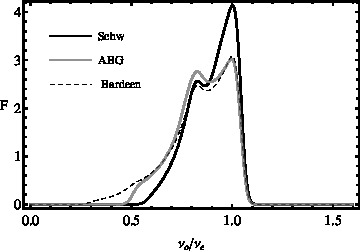}&
			\includegraphics[scale=0.3]{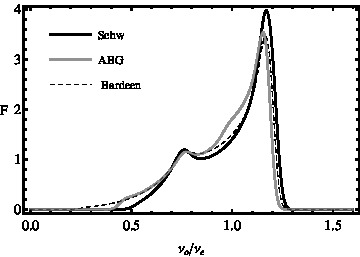}&
			\includegraphics[scale=0.3]{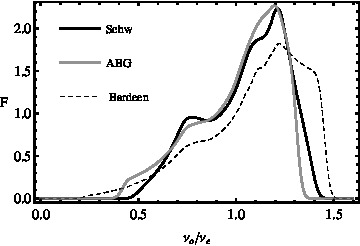}
		\end{tabular}
		\caption{The profile of spectral line generated in the Keplerian disc orbiting the Schwarzchild black hole and the Bardeen and ABG Class IV no-horizon spacetimes. The charge parameter is chosen as $g=1.0$ in both the Bardeen and ABG spacetimes. The observer inclination is $\theta_o=30^\circ$ (top), $60^\circ$ (middle), and $85^\circ$ (bottom). \label{pl_1n0}}
	\end{center}
\end{figure}


\begin{figure}[H]
\begin{center}
\begin{tabular}{ccc}
\includegraphics[scale=0.4]{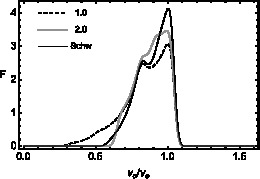}&\includegraphics[scale=0.4]{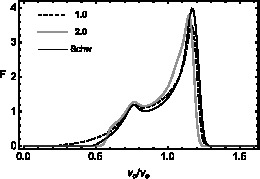}&\includegraphics[scale=0.4]{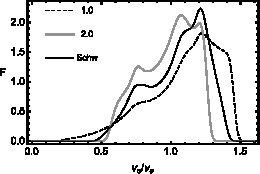}\\
\includegraphics[scale=0.4]{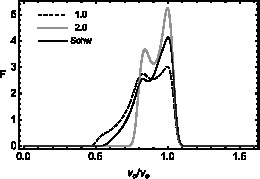}&\includegraphics[scale=0.4]{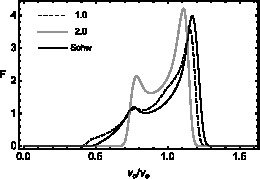}&\includegraphics[scale=0.4]{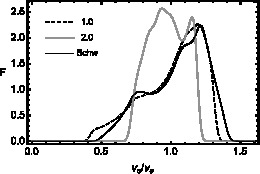}
\end{tabular}
\caption{Profiles of spectral lines of radiation generated in the Keplerian disc orbiting the Bardeen (top) and ABG (bottom) Class IV no-horizon spacetimes with two representative values of the charge parameter $g=1.0$(dashed) and $2.0$(dotted) and three values of inclination $\theta_o=30^\circ$(left), $60^\circ$(middle), and $85^\circ$(right). The black solid curves correspond to the case the profiled line generated in the field of the Schwarzchild black hole.}
\end{center}
\end{figure}

\begin{figure}[H]
	\begin{center}
		\begin{tabular}{ccc}
			\includegraphics[scale=0.4]{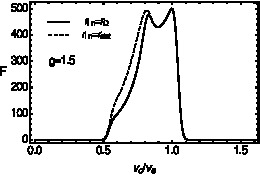}&\includegraphics[scale=0.4]{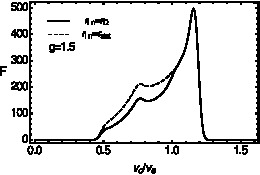}&\includegraphics[scale=0.4]{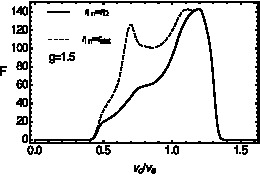}
		\end{tabular}
		\caption{Profiles of spectral line of radiation emitted from Keplerian disk formed in Bardeen metric field with $g=1.5$, spanning between $r_{in}=r_{\Omega}$ (solid), $r_{stat}$ (dashed) resp.  and $r_{out}=20M$. The inclination of the observer is $\theta_o=30^\circ$ (left), $60^\circ$ (center), and $85^\circ$ (right).}
	\end{center}
\end{figure}


For comparison of the profiled lines we use in Figure 20 the common charge parameter $g/m=1$, and we observe a significant relative shift of the profiled lines in the Bardeen and ABG spacetimes and their strong modification in comparison to the Schwarzschild profiled lines. 

For small inclination, the Bardeen and ABG profiled lines are flatter in comparison to the Schwarzschild line, their height is substantially lowered relative to the Schwarzschild case, their frequency range is similar and nearly equal to the Schwarzschild range at the blue-end, but the range is substantially increased at the red-end, being much larger for the Bardeen case in comparison to the ABG case. The ABG profiled line has the red-end peak only slightly lower than the blue-end peak. In the Bardeen spacetime, the difference of the heights of the two peaks is larger than in the ABG spacetime.  

For the mediate inclination, the line height is lowered (slightly more in the Bardeen case) relative to the Schwarzschild case. The frequency range of the Bardeen and ABG lines is shifted relative to the Schwarzschild range to lower values at both ends. At the blue-end the shift is insignificant, being larger for the ABG line, while its is substantial at red-end, being significantly larger for the Bardeen line. 

For large inclination, the height of the Bardeen line is significantly reduced in comparison to the height of the Schwarzschild line, while the height of the ABG profiled line is slightly higher and substantially wider than for the Schwarzschild line. The frequency range of the Bardeen and ABG lines is shifted relative to the range of the Schwarzschild line to lower values at the red-end (the shift being substantially larger in the Bardeen case), and to the lower (higher) values at the blue-end for the ABG (Bardeen) line. 

From Figure 21 we can conclude that for both the Bardeen and ABG Class IV no-horizon spacetimes, and for all inclination angles of the Keplerian disc, increasing charge parameter causes substantial modifications of the profiled lines and substantial restriction of their frequency range. In all cases, the changes have even qualitative character for the substantial increasing of the charge parameter to the value of $g/m=2$. Then the frequency range of the Bardeen and ABG spacetimes is for all inclinations located inside of the frequency range of the Schwarzschild line. The changes are clearly most profound for large inclination angle when even multipeak profiled lines could be created due to the occurrence of the ghost images. For small inclination, the modifications due to increasing of the charge parameter $g/m$ tend to creation of one peak line in the Bardeen spacetimes, while a narrow line profile with clearly distinguished two peaks is generated in the ABG spacetimes, where this tendency survives for the profiled lines created under mediate inclination of the Keplerian disc. 

In Figure 22 we represent for the Bardeen no-horizon spacetime with $g/m=1.5$ the input of radiation occuring from the Keplerian disc located under the radius $r_\Omega$ to the profiled spectral lines. We demonstrate that such an input is relevant and observable especially for large inclination angles. The photons coming from $r<r_\Omega$ significantly amplify the red wing and central region of the profiled line, as they origin in the slowly rotating part of the disc where Doppler shift is suppressed. In the ABG spacetimes the same effects are relevant.  

The profiled lines created by Keplerian discs orbiting in the Bardeen and ABG Class IV spacetimes give clearly measurable signatures. For sufficiently large charge parameters ($g/m>1$) the differences between the Bardeen and ABG profiled lines can be even of qualitative character for all considered inclination angles. Note that such kind of behavior does not occur in the naked singularity Kehagias-Sfetsos spacetimes of modified Ho\v{r}ava quantum gravity \cite{Stu-Sche:2014:CLAQG:}. 

\subsection{Contribution of the ghost images} 

Finally, we shortly illustrate the role of the radiation incoming from the ghost images to the profile of the spectral lines. Recall that in the no-horizon spacetimes the ghost images are created by a small part of the Keplerian disk near its inner edge,  momentarily located just behind the centre of the geometry while being on the line connecting the centre and the observer. The ghost images are created by photons with small impact parameters following  almost radial geodesics crossing the nearly flat central region of the no-horizon spacetime -- this is the reason why the inclination angle of the observer has to be very large \cite{Sche-Stu:2015:JCAP:}. In the naked singularity spacetimes (of RN or KS type) the photons are repulsed by the central region of the spacetime and the ghost images have different character as demonstrated in \cite{Sche-Stu:2015:JCAP:}.  

We construct in the relevant cases of the Bardeen spacetimes (with charge parameter large enough to enable creation of the ghost images) and for large inclination angle ($\theta_{o}=85^{\circ}$) necessary for creation of sufficiently large ghost images \cite{Sche-Stu:2015:JCAP:} the profiled lines for the total incoming radiation, and those without the contribution of the ghost images. 

\begin{table}[H]
	\begin{center}
		\caption{Ratio of bolometric fluxes calculated for the three representative values of the Bardeen spacetime parameter $g/m$.\label{flux_bol}}
		\begin{tabular}{|c|ccc|}
			\hline
			$g$ & $1.5$ & $2.0$ & $2.5$\\
			\hline
			$F_{NoG}/F_{G}$ & $0.985$ & $0.973$ & $0.947$\\
			\hline
		\end{tabular}	
	\end{center}
\end{table}

\begin{figure}[H]
	\begin{center}
		\begin{tabular}{ccc}
			\includegraphics[scale=0.3]{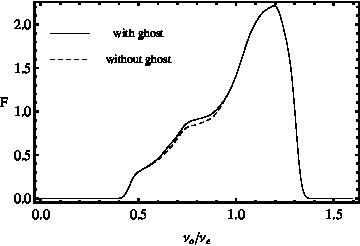}&
			\includegraphics[scale=0.3]{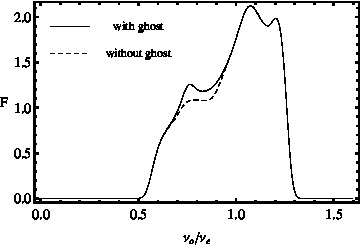}&
			\includegraphics[scale=0.3]{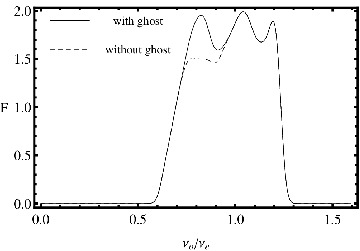}
		\end{tabular}
		\caption{Comparison of the profiles of spectral lines constructed in two situations. First, the line is formed from the whole primary disk image. Second, the line is constructed for the primary disk image with ghost image region removed. The inclination of observer is $\theta_o=85^\circ$ and the spacetime parameter $g=1.5$, $2.0$, and $2.5$ (from left to right).\label{ghost} }
	\end{center}
\end{figure} 

We demonstrate in Figure 23 that the role of the ghost images cannot be abandoned, and in some cases could be quite relevant. We can see that the role of the ghost images increases with increasing charge parameter, and its signature is clearly reflected in the innermost peak created in the multi-peaked profiled line. The quantitative reflection of the increasing role of the ghost image with the dimensionless charge parameter increasing is reflected by the ratio of the bolometric luminosities represented in Table 3. 

\section{Conclusions}

In the present paper we complete the results of our previous extended study of the optical phenomena in the regular no-horizon spacetimes and naked-singularity Reissner-N\"{o}rdstr\"{o}m spacetimes where attention was concentrated on the so called ghost images \cite{Sche-Stu:2015:JCAP:}. 

We have constructed the profiled spectral lines in the field of regular black holes of the Bardeen and ABG type, and their "no-horizon" counterparts, in dependence on the dimensionless charge parameter of the spacetime $g/m$ reflecting fully properties of the spacetime. The influence of the electromagnetic field related to the Bardeen and ABG spacetimes has not be considered. 

The no-horizon spacetimes have their astrophysical properties very similar to those found for the first time in the case of the spherically symmetric Reissner-Nordstron-de Sitter naked singularity spacetimes \cite{Stu-Hle:2002:ActaPhysSlov:}, and recently discussed for the naked singularity spacetimes of the Reissner-Nordstrom type related to the Einstein gravity \cite{Pug-Que-Ruf:2011:PHYSR4:}, or to the spherically symmetric braneworld black holes \cite{Stu-Kot:2009:GenRelGrav:}, and to the Kehagias-Sfetsos naked singularity spacetimes of the Ho\v{r}ava quantum gravity \cite{Stu-Sche-Abd:2014:PHYSR4:,Vie-etal:2014:PHYSR4:,Stu-Sche:2014:CLAQG:}. For the no-horizon Bardeen and ABG spacetimes, three different regimes of circular geodesics occur in dependence on the parameter $g/m$ \cite{Stu-Sche:2015:IJMPD:}. For all three regimes, a "static sphere" representing the innermost limit on the existence of circular geodesics exists due to some antigravity effects arising near the coordinate origin - therefore, we call it also "antigravity sphere". For large values of $g/m$, stable circular geodesics exist at all radii above the static sphere. For intermediate values of $g/m$, two regions of stable circular geodesics exist, being separated by a region of unstable circular geodesics. For lowest values of $g/m$ compatible with the no-horizon Bardeen and ABG solutions, an inner region of stable circular geodesics, beginning at the static radius, is limited from above by a stable photon circular geodesics, and an outer region, beginning at ISCO, extends up to infinity. In all cases we have considered as the radiating Keplerian disc the region of the stable circular geodesics where angular frequency of the orbital motion increases with decreasing radius, and the MRI instability governing the Keplerian accretion can work as discussed in \cite{Stu-Sche:2014:CLAQG:}. 

It has been demonstrated that in the case of the regular black hole spacetimes the profiled spectral lines are of the same character as in the Schwarzschild spacetime and the quantitative difference related to the frequency range of the profiled spectral line is generally larger for the ABG black holes than for the Bardeen ones. 

Bardeen and ABG no-horizon spacetimes of all these three types of the character of the circular geodesic motion can be distinguished by the behavior of the profiled spectral lines, if the inclination angle to the disc is known. Large differences in the shape and the frequency range of the profiled spectral lines has been demonstrated in the case of the Bardeen and ABG no-horizon spacetimes with large values of the spacetime charge parameter $g/m>1$. These differences enable also a clear distinguishing of the Bardeen and ABG no-horizon spacetimes. Moreover, there exist also a clear distinction of the Bardeen and ABG profiled lines, and the profiled lines generated in the Kehagias-Sfetsos naked singularity spacetimes of the modified Ho\v{r}ava quantum gravity. We expect that recent observational techniques could enable to distinct the regular no-horizon and the naked singularity spacetimes. 

\section*{Acknowledgements}
The authors acknowledge institutional support of the Faculty of Philosophy and Science of the Silesian University at Opava, and the Albert Einstein Centre for Gravitation and Astrophysics supported by the~Czech Science Foundation Grant No. 14-37086G.

\end{document}